\documentclass[prb,twocolumn,aps,showpacs]{revtex4}

\usepackage{graphicx}

\def\B#1{\bm{#1}}

\usepackage{bm}
\usepackage{amssymb}
\usepackage{amsfonts}
\usepackage{amsmath}
\usepackage{graphicx}    
\usepackage[dvips]{color} 

\definecolor{g-blue}{rgb}{0.83,0.95,1}
\definecolor{Blue}{rgb}{0.5,0.5,1}
\definecolor{DarkBlue}{rgb}{0.00,0.00,0.58}
\definecolor{g-yellow}{rgb}{1,1,0.7}
\definecolor{g-green}{rgb}{0.9,1,0.9}
\definecolor{green}{rgb}{0,0.6,0}
\definecolor{Green}{rgb}{0,0.4,0}
\definecolor{cyan}{rgb}{0,0.7,0.7}
\definecolor{black}{rgb}{0,0,0}
\definecolor{grey}{rgb}{0.4 ,0.4 ,0.4 }

\def\blue#1{\textcolor{blue}{#1}}

\renewcommand{\sb}[1]{_{\text {#1}}}  
\renewcommand{\sp}[1]{^{\text {#1}}}  
\def\Sb#1{_{\scriptscriptstyle\rm{#1}}}
 \def\= {\equiv}

\newcommand{\C}[1]{{\mathcal{#1}}}    
\def \ed {\end{document}}
\def\Fbox#1{\vskip1ex\hbox to 8.5cm{\hfil\fboxsep0.3cm\fbox{%
  \parbox{8.0cm}{#1}}\hfil}\vskip1ex\noindent}  
\def\<{\left\langle}    \def\>{\right\rangle}
\def\({\left(}          \def\){\right)}
\def \[ {\left [} \def \] {\right ]}
\def\~{\widetilde}

\begin{document}
\title{Kelvin waves and the decay of quantum superfluid turbulence}
\author{Luiza Kondaurova$^{1,2}$, Victor L'vov$^{2}$,  Anna Pomyalov$^{2}$ and Itamar Procaccia$^{2}$}
\today
\affiliation{$^1$ Institute of Thermophysics, Novosibirsk 630090, Russia\\
$^2$ Department of Chemical Physics, The Weizmann Institute of Science, Rehovot
76100, Israel
}

\begin{abstract}
We present a comprehensive statistical study of free decay of the quantized vortex tangle in superfluid $^4$He at low and ultra-low temperatures, $0\leqslant T \leqslant  1.1\,$K. Using high resolution vortex filament simulations with full Biot-Savart vortex dynamics,   we show that for ultra-low temperatures $T\lesssim 0.5 \,$K, when the mutual friction parameters $\alpha\simeq \alpha' < 10^{-5}$,  the vortex reconnections excite Kelvin waves with wave lengths $\lambda$ of the order of the inter-vortex distance $\ell$.  These excitations cascade down to the resolution scale $\Delta\xi$  which in our simulations is of the order $\Delta \xi\sim \ell/100$. At this scale the Kelvin waves are  numerically damped  by a line-smoothing procedure, that is supposed to mimic the dissipation of Kelvin waves by  phonon  and roton emission at the scale of the vortex core. We show that the Kelvin waves  cascade  is statistically important: the shortest available Kelvin waves at the end of the cascade determine the mean vortex line curvature $S$, giving $S \gtrsim 30 /\ell$ and play  major role in the tangle decay at ultra-low temperatures below $0.6\,$K. The found dependence of $\ell S$ on the resolution scale $\Delta \xi$ agrees with the L'vov-Nazarenko energy spectrum of weakly-interacting Kelvin waves, $E\Sb{LN}\propto k^{-5/3}$ rather than the spectrum $E\Sb{LN}\propto k^{-1}$, suggested by Vinen for turbulence of Kelvin waves with large amplitudes. We also show that  already at  $T=0.8\,$K, when $\alpha$ and $\alpha'$ are still very low, $\alpha\simeq \alpha'<10^{-3}$, the Kelvin wave cascade is fully damped, vortex lines are very smooth, $S \simeq  2 /\ell$ and the tangle decay is predominantly caused by the mutual friction.

\end{abstract}
\maketitle

\section{Introduction}
\label{intro}
Below the Bose-Einstein condensation temperature $T_\lambda\approx 2.18\,$K liquid $^4$He becomes a quantum superfluid, demonstrating inviscid behavior\,\cite{1}. Aside from  the lack of viscosity, the vorticity in $^4$He  is constrained  to vortex-line singularities of fixed circulation $\kappa= h/M$, where $h$ is Planck's constant and $M$ is the mass of the $^4$He atom. These vortex lines have a core radius $a_0\approx 10^{-8}\,$cm, compatible with  the inter-atomic distance.
In generic turbulent states these vortex lines appear as a complex tangle and we denote the typical inter-vortex distance as $\ell$.

There is a growing consensus\,\cite{V00,VN02,Vin10,2,BLR14} that at large scales $R\gg \ell$, superfluid turbulence exhibits statistical properties that are akin to classical turbulence. In particular, energy cascades in classical and superfluid turbulence are similar up to the inter-vortex scale $\ell$  where the discreteness of the vorticity  becomes important. It is also believed that in the zero temperature limit the energy is further transferred down scales by interacting Kelvin waves \cite{S95,KS1,AT00,VTM03,LN1} which are helical perturbation of the individual vortex lines\,\cite{Ke1880}. Finally at the core radius scale the energy is radiated away by phonons and rotons\,\cite{V00,V01}.

The direct observation of Kelvin waves which are excited by vortex reconnections was recently achieved by visualizing the motion of submicron particles dispersed in superfluid  $^4$He cf. Ref. \cite{3}. Nevertheless, as far as we are aware, there are as yet no experimental observation of the Kelvin wave cascade in superfluid $^4$He, cf. Ref. \cite{VTM03,EGGHKLS07,SS12,BSS14}. Therefore at present the confirmation of the statistical importance  of Kelvin waves in the evolution of superfluid turbulence should be achieved with computer simulations.  This is the main goal of this paper.

Computer simulations of quantum vortex tangle dynamics were pioneered in Ref. \cite{Schwarz88} with much following
research cf.~Refs.\cite{AT00,VTM03,BSS14,Schwarz88,H13,HB14,BLR14,KLPP14,KS2,11,12,KVSB01,TAN-2000,27,KN-12} and references therein. Some studies focused on vortex line evolution starting from particular and simple initial conditions, see e.g.  Ref\,\cite{AT00}, where a vortex ring approaches a rectilinear vortex or Ref.\,\cite{KVSB01} in which four vortex rings collide. In these and similar simulations Kelvin waves are excited by vortex reconnections and evidence was found for the development of  the direct (from large to small scales)  cascade of Kelvin waves. Unfortunately the numerical resolution was not sufficient to distinguish between available theoretical predictions for the energy spectra of Kelvin waves $E(k)$.

There exist three different predictions for this spectrum. The first was proposed by
Vinen\ \cite{VN02}, and is expected to hold when the amplitudes of the Kelvin waves are large (strong Kelvin wave turbulence):
 \begin{subequations}\label{spectra}
 \begin{equation}\label{V}
 E\Sb V(k)= E \big / k \ .
  \end{equation}
  Two other prediction were offered for the case of weak Kelvin wave turbulence (small amplitude waves). The first
  was offered by  Kozik  and Svitunov\cite{KS1}  (KS)  under the assumption of the locality of interactions
 \begin{equation}\label{KS}
 E\Sb {KS}(k)= E \big / \big ( \ell^{2/5} k^{7/5}\big) \ .
 \end{equation}
 The second was predicted by L'vov and Nazarenko\cite{LN1} (LN)   who argued that the interaction of Kelvin waves is not local:
 \begin{equation}\label{LN}
 E\Sb {LN}(k)= E \big / \big ( \ell^{2/3} k^{5/3}\big)\ .
 \end{equation}
 \end{subequations}
 Here,   $E\sim \int_{1/\ell}^{1/a_0}E(k)dk$ is the  energy density of Kelvin waves per unite vortex line length, normalized by the $^4$He density, up to dimensionless numerical constants. Equations\,\eqref{KS} and  \eqref{LN} should describe exactly the same physical situations and definitely are in contradiction, leading to some debate\cite{72,73,74,75,76,77,78,BB11}. In fact, recent simulations of the Gross-Pitaevskii equation  by Krstulovic\,\cite{Krs12} support the spectrum\,\eqref{LN}. In the most recent vortex filament simulations by Baggaley and Laurie\,\cite{BL13} a remarkable agreement with the LN spectrum\,\eqref{LN} was reported, including the dimensionless coefficient $C\Sb{LN}\sp{num}\approx 0.308$, very close to the analytic prediction of 0.304 (cf Ref. \ \cite{79}).

While it appears that Eq.\,(\ref{LN}) takes preference on Eq.\,(\ref{KS}) for weak Kelvin wave turbulence, the crucial next question is whether Kelvin waves turbulence is indeed weak in realistic experimental situations, say in the decay of counterflow turbulence. If Kelvin wave turbulence happens to be strong, then the Vinen spectrum is likely to be observed.

This  paper aims to clarify, at least partially, these and related questions with the help of high-resolution numerical simulation of the decay of counterflow turbulence in $^4$He at low temperatures $ T \leq 1.1\,$K, including the zero temperature limit. Details of the numerical procedure are described in the  Sec.\,\ref{s:num}, the results and their discussion are given in Sec.\,\ref{s:results}.
In particular, Sec.\,\ref{ss:VLD} is devoted to the time evolution of the vortex line density and  Sec.\,\ref{ss:recon} to  the reconnection rate in the free decaying tangle. In the analytical Sec.\,\ref{ss:curv} we present a relation between the spectrum of Kelvin waves and the root-mean-square (rms)  vortex curvature.
    Section\,\ref{ss:curvA} presents our numerical results for  the evolution of the  rms  vortex-line curvature.  The analytical and numerical results of Secs.\,\ref{ss:curv} and \ref{ss:curvA} allow us to estimate in Sec.\,\ref{ss:est} the upper cutoff of the Kelvin wave spectrum.  Section\,\ref{ss:res0lution} discusses the details  of the tangle decay at different spatial resolutions  and reaches
     the important conclusions that in the zero temperature limit vortex tangles exhibit a weak Kelvin wave turbulence. A short summary of the results is given in the end of the paper.

\section{\label{s:num}Numerical procedure}
Our simulations are aimed at studying the decay of a vortex tangle using the vortex filament method\cite{Schwarz88}. The details of the implementation were described at length in Ref.\cite{KLPP14}.
The simulations were carried out in the cubic box $H=0.1$ cm for temperatures $T=0\,, \   0.43\,, \ 0.55\,, \ 0.8\,,\   0.9\,$K, and  $1.1\,$K. The evolution of a point $\bm s$ on the vortex line is described by
 \begin{subequations}\label{BS}
\begin{eqnarray}\label{eq:s_Vel}
\frac{d{\bm s}}{dt} &=&{\bm V}_{\rm s}+{\bm V}_{\rm si}-\alpha  {\bm s}'\times  {\bm V}_{\rm si}+\alpha' {\bm s}' \times \Big[{\bm s}' \times {\bm V}_{\rm si}\Big]  .
\end{eqnarray}
This equation was solved using the Biot-Savart representation of the velocity, defined by the entire vortex tangle
\begin{eqnarray}\label{BSE}
{\bm V}_{\rm si}({\bm s}) =  \frac{\kappa}{4\pi}\int_{ \C C}
 \frac{({\bm s_1}-{\bm s})\cdot  d s_1}{|{\bm s_1}-{\bm s}|^3}+{\bm v}_{\rm bc} \ .
\end{eqnarray}\end{subequations}
Here the vortex line is presented in a parametric form $ \bm s(\xi,t) $, where $\xi$
 is an arclength,  $ \bm  s'  =d \bm  s/d\xi $  is
a local direction of the vortex line,
 $t$ is the time and the integral is taken over the entire vortex tangle configuration $\C C$. The influence of the boundary conditions is accounted for in  ${\bm v}_{\rm bc}$.  ${\bm V}_{\rm s}$ is the macroscopic super-fluid velocity
component. In the reference frame co-moving with the superfluid component, $V_{\rm s}=0$.

The mutual friction parameters $\alpha(T)$ and $\alpha'(T)$ are presented in Table\,\ref{t:1}.   For $T=0.8$, $0.9$ and $1.1$\,K they are taken from Refs. \cite{BDV83,DB98} by extrapolating the data for $B$ and $B'$ from Figs. 1 and 2 in  Ref.\cite{BDV83} and using the relations between $B,B'$ and  $\alpha, \alpha'$  and the data for $\rho\sb n$ and $\rho$ from Ref.\cite{DB98}. For lower temperatures we used an approximate equation from Ref.\,\cite{Rev09}
\begin{equation} \label{alpha}
\alpha\simeq \frac{25.3}{\sqrt T} \exp\Big ( - \frac{8.5}{T}\Big ) + 5.78 \,\big(0.1 \, T\big)^5\,,
\end{equation}
in which $T$ is in Kelvin. The first term, dominating for $T>0.5\,$K, originates from the roton scattering, discussed in details by Samuels and Donnelly\,\cite{SD90}; the second one, taken from Iordanskii's paper\,\cite{Ior66}, describes the contribution of the phonon scattering.  Without a corresponding equation for $\alpha'$ in the available literature we took for $T=0.43$ and 0.55\,K $\alpha=\alpha'$ according to the trend at the higher temperature data, see Table\,\ref{t:1}. Some researches used to neglect the  effect of $\alpha'$ at low temperatures, arguing that $\alpha'\ll 1$. Indeed, $\alpha'$ is small, e.g. $\alpha'\approx 0.5\cdot  10^{-3}$ at $T=0.8\,$K. Nevertheless, as we show in Appendix\,A, even very small $\alpha'$ (but comparable with $\alpha$) affects the dynamics of the vortex tangle quite strongly.

 In our simulations periodic conditions were used in all directions.
In vortex filament simulations the reconnections between vortex lines are introduced artificially. The influence of three different popular reconnection criteria\,\cite{AFT10,Samuels92,KAN08} on the results of such simulations  were studied in details in\,\cite{KLPP14}. The first criterion  requires only that the two filament points come closer than a predefined distance to trigger reconnections. The second criterion requires reduction of the line length in addition to the geometric proximity. The last criterion is the most strict one and requires that the reconnecting line segments cross in space during the next time step.  It was shown that two criteria \cite{Samuels92,KAN08}  give similar results for all the calculated properties of the vortex tangle, while the criterion\cite{AFT10}  gave rise to a very large number of reconnections and large number of small loops that have to be removed algorithmically to ensure stability of the simulations. For simulations of the tangle decay at very low temperature the role of reconnections becomes very important. Therefore we have chosen the most efficient numerical reconnection procedure that gives reliable reconnection rate, the criterion \cite{Samuels92}.
The reconnections were performed when two line points become closer that the  initial line resolution distance $\Delta \xi_{\rm init}$, (that was chosen to be $\Delta \xi_{\rm init}=5 \cdot  10^{-4}$ cm)  provided the total line length is reduced upon reconnection.

 The  sensitivity of the results  to the resolution  was clarified by simulating with  smaller and larger values of $\Delta \xi_{\rm init}$ as discussed below.  Points along the vortex line were removed or added during evolution to keep the resolution $\Delta \xi$ inside the interval $\Delta \xi_{\rm init}/1.5\le \Delta \xi\le 1.5\Delta \xi_{\rm init}$.

 This smoothing procedure is required to ensure the accuracy of the calculation of the Biot-Savart integral besides serving as a numerical filter for the high-frequency oscillations of the vortex line,corresponding to the short Kelvin waves with the wave length $\lambda\sb{min}$ about $2\Delta \xi$. Removal of the short Kelvin waves serves as their effective damping mechanism at the smallest scale edge of the \emph{direct} energy cascade achieved in the numerical simulations  $\lambda\sb{min}\simeq 2 \Delta\xi$. We think that this  numerical cutoff in the wave number space does not affect the
actual direct cascade in the inertial  interval of wave-lengths $\lambda\gg \lambda\sb{min}$ and therefore leads to the same energy spectrum  as an  actual dissipation of Kelvin waves at the vortex core radius by radiation of phonons and rotons\,\cite{V00,V01}.

  A possible worry may be based on the existence of the \emph{inverse}  (from small to large wave-lengthes) cascade of the action of Kelvin waves $n(k)$ Ref.\,\cite{Naz06}, related to their energy density $E(k)$ and frequency $\omega(k)$ as follows: $n(k)=E(k)/\omega(k)$, see footnote\,\footnote{The wave action $n(k)$ can be considered as a classical limit of the occupation numbers $N(k)$ in the limit $N(k)\gg 1$: $n(k)\to h N(k)/2 \pi$, where   $h$ is the Plank constant.}. The inverse cascade  may  potentially affect  scales $\lambda\gg \lambda\sb{min}$. However in our case there is no source of Kelvin waves  action $n(k)$ at $\lambda\simeq \lambda \sb{min}$, only their dissipation,  and thus the inverse cascade is not excited.

\begin{table}
\begin{tabular}{||c|| c| c| c| c| c| c||}
  \hline \hline
  ~$T\,,\,$K~ & ~~0~~ &   0.43&   0.55  & ~0.8~ & ~0.9~ & ~1.1~ \\ \hline
  $\alpha\cdot  10^3$ &  0 &  $0.001$ &   $0.01$& $0.649  $ & ~$2.3  $ & ~$11.2  $  \\
   $\alpha'\cdot  10^3$ &  0 &  $0.001$ &   $0.01$ &   $~\,0.4635  $ &\, $1.4 $ & $~~5.6  $\\
  \hline \hline
\end{tabular}
\caption{\label{t:1} Parameters of mutual friction at different temperatures, used in numerical simulations.  }
\end{table}
\begin{figure*}
\begin{tabular}{|c|c|c|}
  \hline
 A & B & C\\
\includegraphics[width=5.8 cm]{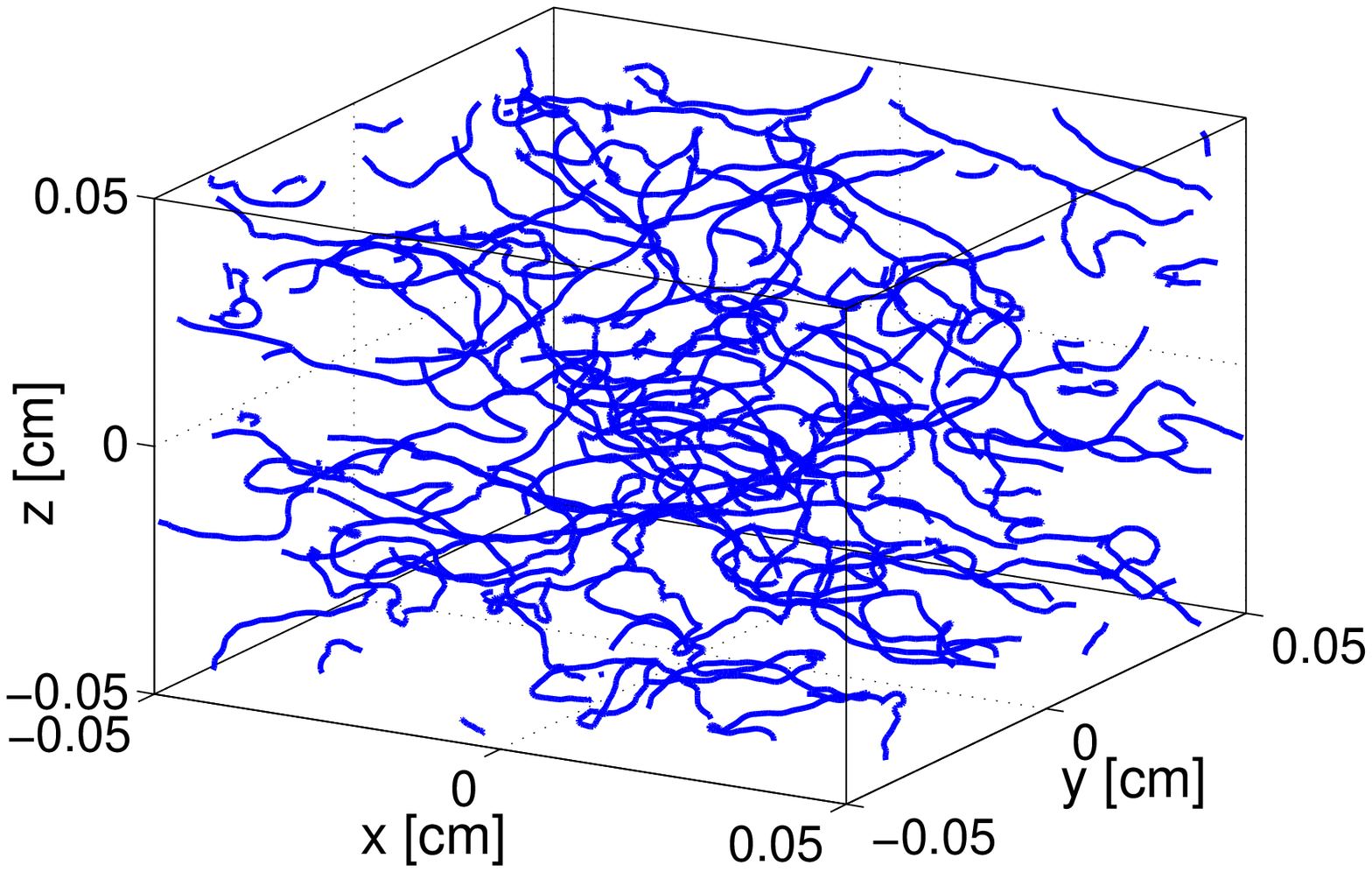}   &
\includegraphics[width=5.8 cm]{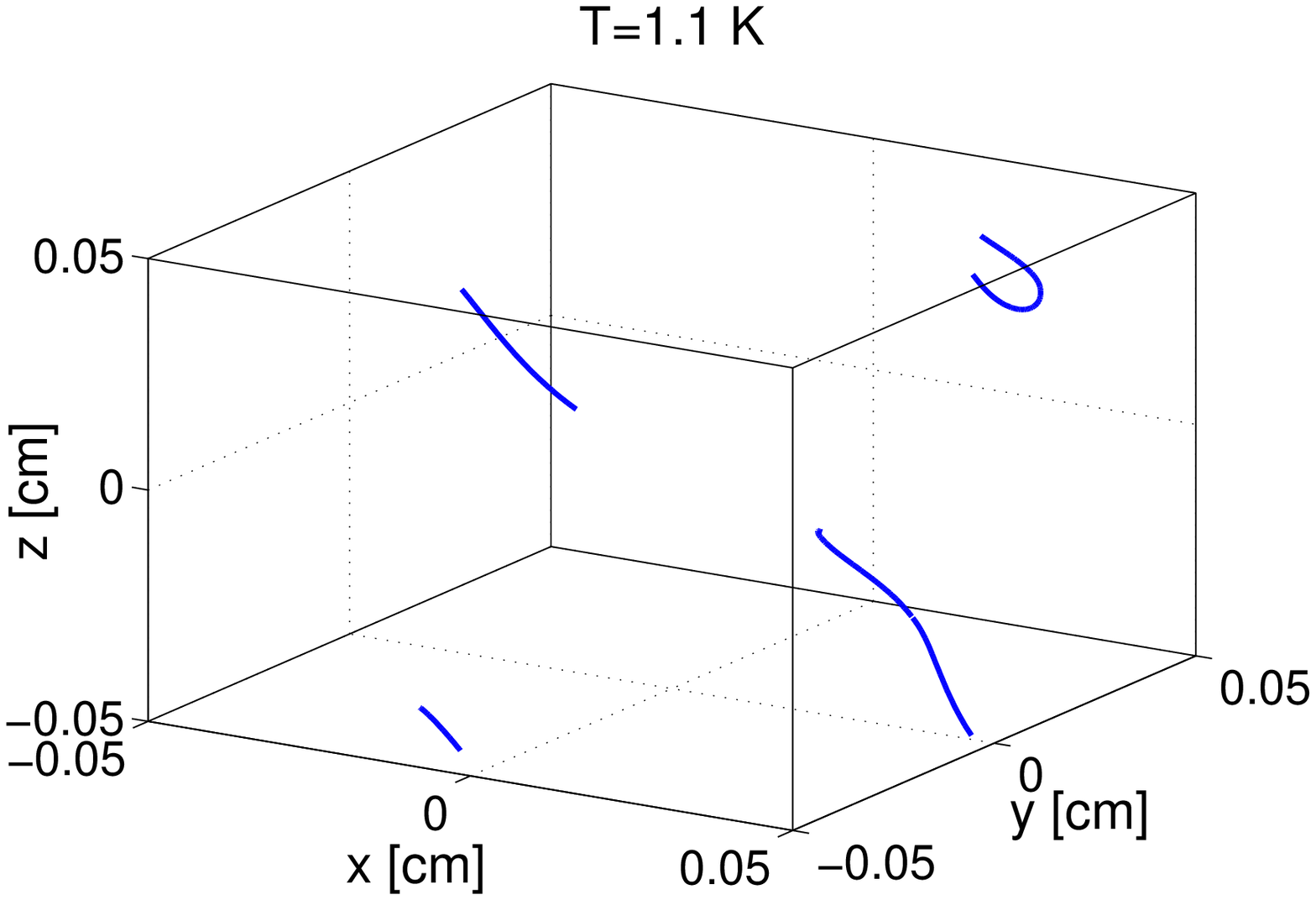} &
 \includegraphics[width=5.8 cm]{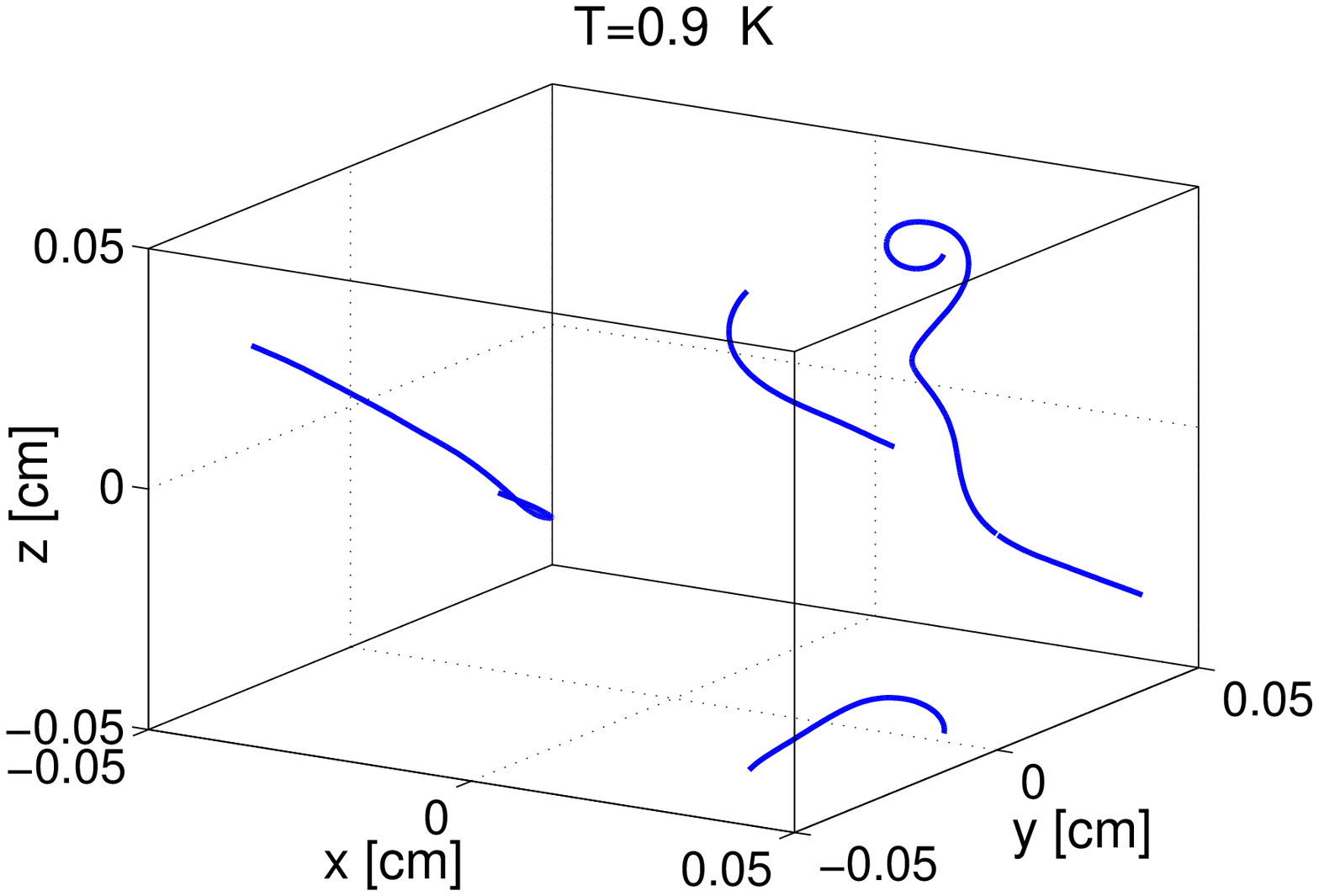}
\\ \hline
D & E & F \\
 \includegraphics[width=5.8 cm]{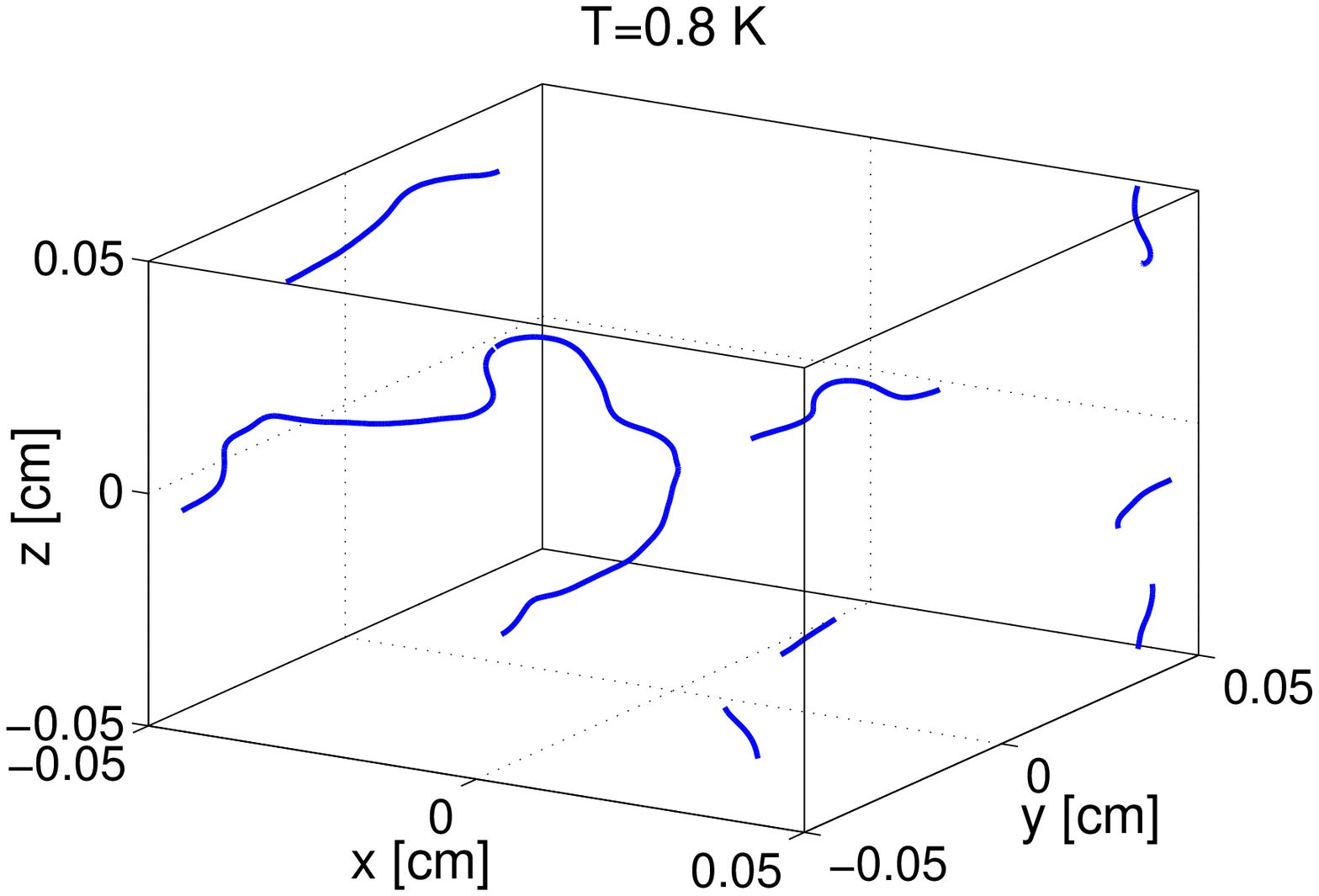}&
\includegraphics[width=5.8 cm]{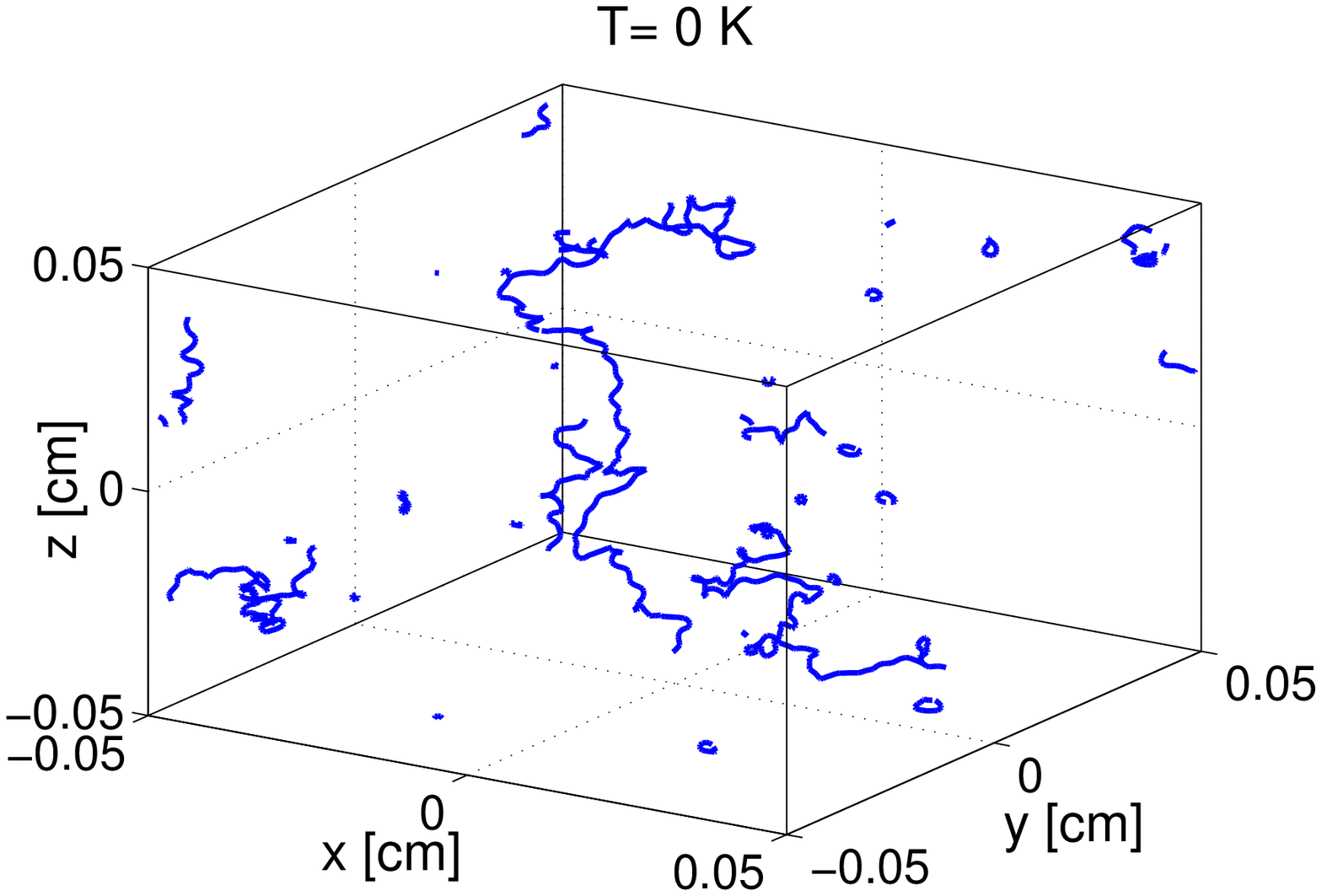} &
 \includegraphics[width=5.8 cm]{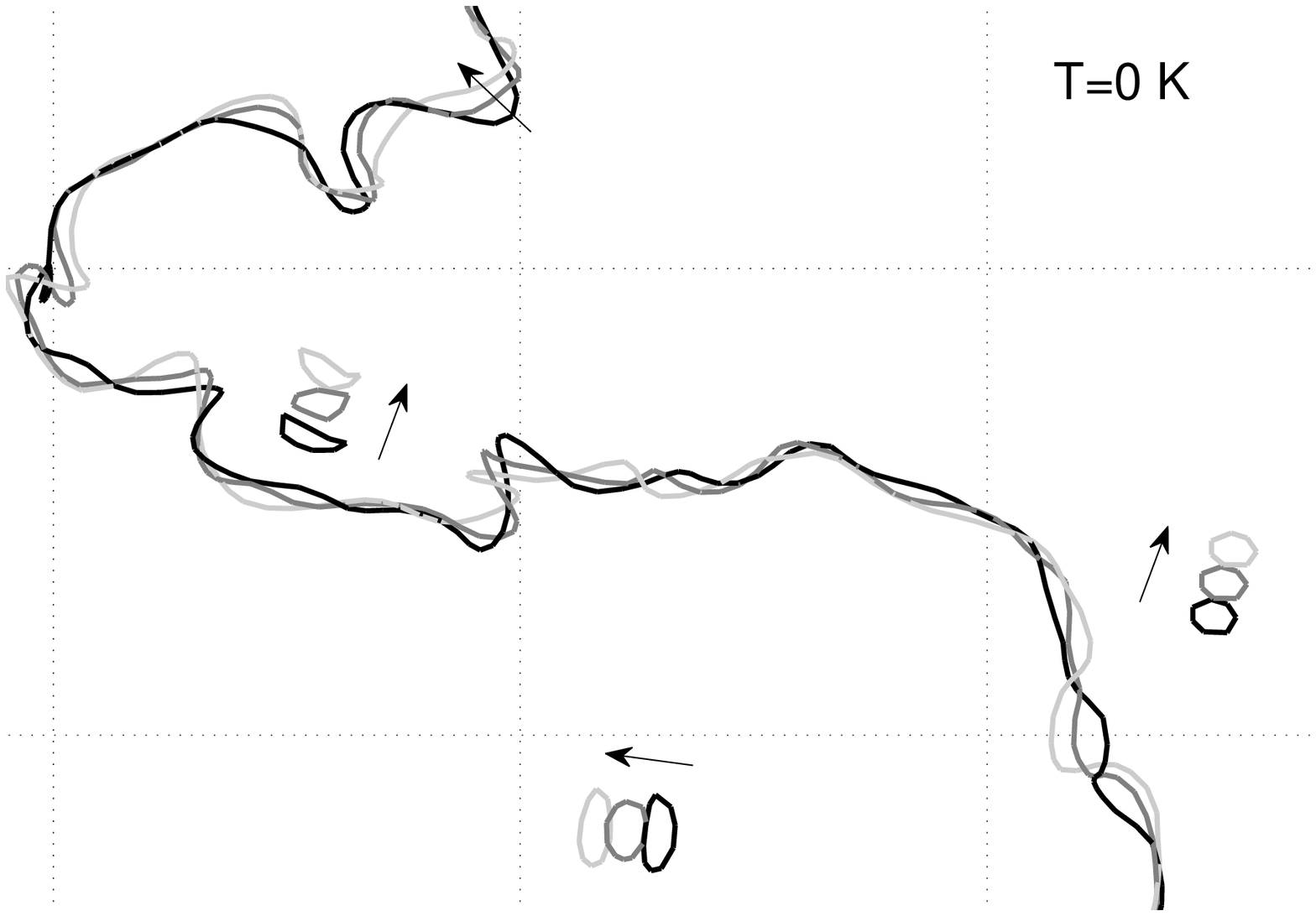}\\
  \hline
\end{tabular}
\caption{\label{f:1}Color online. \emph{Panel A.}  An example of the initial configuration, $t=0\,$s,
used in the simulations at all the temperatures shown in Panels B-F.  \emph{Panels B-E}.
Configurations obtained at $t=50\,$s at different temperatures.   \emph{Panel~F}.
The fragment  of the tangle configuration, shown in  Panel E  ($T=0$) at three successive times separated by $ 5 \cdot  10^{-4}$ s. The black lines correspond to the earliest time $t=50\,$s, the light grey to the latest time. The arrows indicate  the direction of the line movement.
}
\end{figure*}
We also remove  small loops and loop fragments with three or less line segments. This  serves as an additional dissipation mechanism at wave-lengthes  $\lambda\simeq \lambda\sb{min}$.

 We use the fourth  order Runge Kutta method for the time marching with the time step $\Delta t=5.6 \cdot  10^{-5}$s. Initial conditions
 were prepared by running a steady state counterflow simulations at $T=1.9\,$K. Taking 32 different realization in this run we
 constructed an ensemble consisting of 32
well developed vortex tangle configurations with a similar vortex line density (VLD) of about ${\C L}_0=\C L(t=0)\simeq  6\cdot  10^3$ cm$^{-2}$.
Here the VLD $\C L_j(t)=L_j/\Omega$ of a given vortex tangle (indexed by $j$) is defined    as usual\,\cite{Schwarz88} via its total length
\begin{equation}\label{VLD}
 L_j =  \int _{\C C_j} d\xi\,,
\end{equation}
and  the sample volume $\Omega$.
The integral\,\eqref{VLD} is performed  over the vortex configuration $\C C_j$ in the $j$-tangle, i.e. along all vortices.
For any object $\Psi_j$, found on the particular $j$-configuration $\C C_j$, we define the ensemble average in the usual way:
\begin{equation}\label{ens}
\< \Psi \> \= \frac 1 N \sum_{j=1}^N \Psi_j     \ .
\end{equation}
  An example of an initial vortex configuration is shown in Fig.\,\ref{f:1}A.  With the chosen initial vortex line density the mean initial inter-vortex distance $\ell_0\= 1/ \sqrt{\< \C L_0\> }\approx 0.013$ cm,  which is about 26 times larger than the mean space resolution $\< \Delta \xi \> \approx 5 \cdot  10^{-4}\,$cm.

For this relatively low initial vortex line density it was sufficient to account (in the calculation of the Biot-Savart integral) for the tangle configuration within the main computational domain only. More details on this technical issue can be found in Ref.~\cite{KLPP14}.

The initial vortex configurations were allowed to decay according to
Eq.\,\eqref{eq:s_Vel} with the parameters $\alpha$ and $\alpha'$ corresponding to different temperatures (see Table \,\ref{t:1}). The
 simulation was terminated either when the vortex line density dropped to a background level of about ${\C L}\sb{bg}=100$ cm$^{-2}$ or when a predefined decay time for the given temperature has elapsed.  The value of ${\C L}\sb{bg}$ corresponds, e.g., to a configuration with one straight vortex line over the entire computational domain and parallel to its edge.  The results for each temperature were obtained by the ensemble averaging over the 32 initial configurations.  At the final stage of the simulations the mean intervortex distance  $\ell\sb{bg }\= 1/ \sqrt{\C L\sb{bg}}=0.1\,$cm  is about 200 times larger than  the mean space resolution  $\< \Delta \xi \> $.

\begin{figure*}
\begin{tabular}{|c|c|c|}
  \hline
 A & B & C \\
\includegraphics[width=5.8 cm]{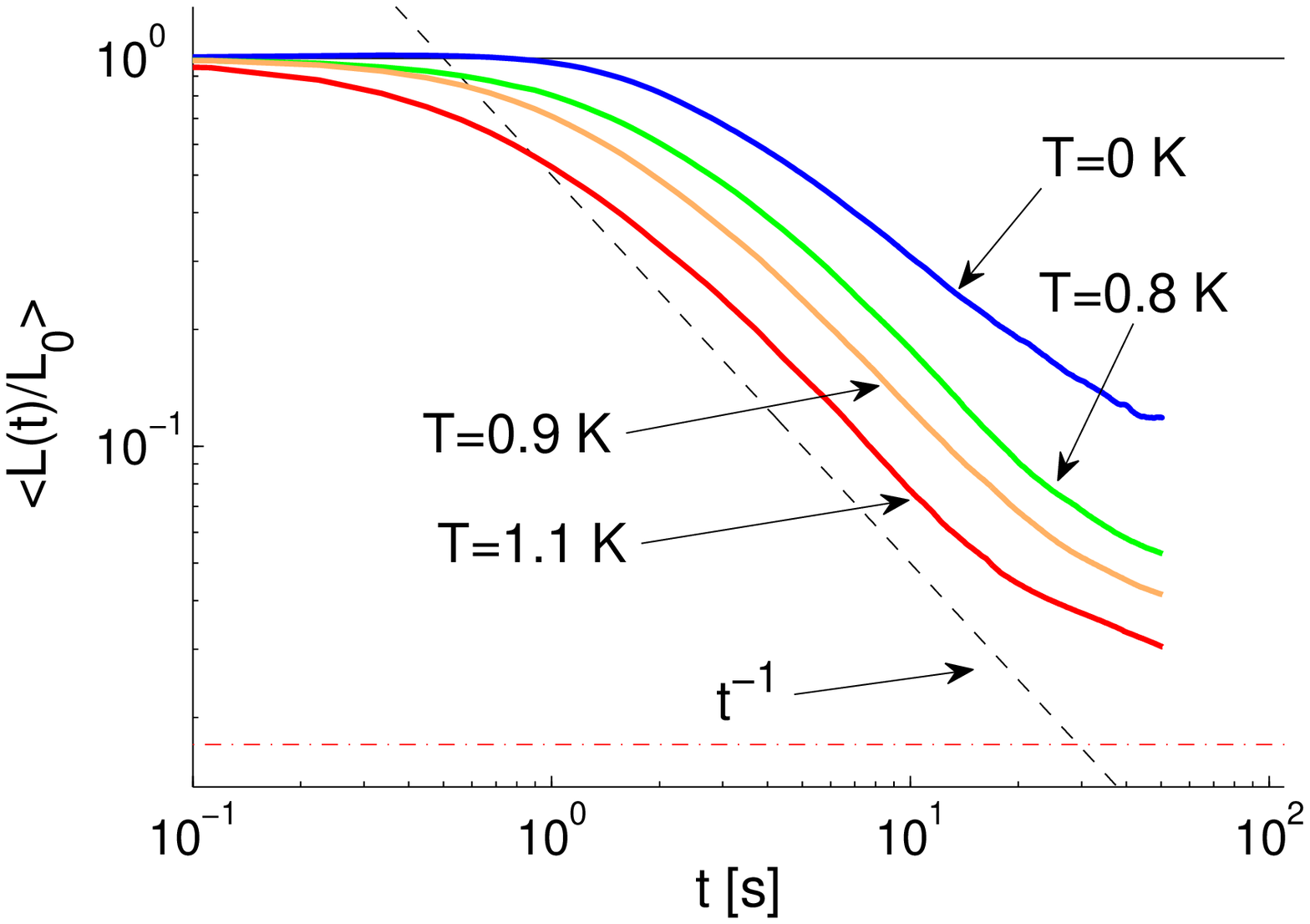}  &
   \includegraphics[width=5.8 cm]{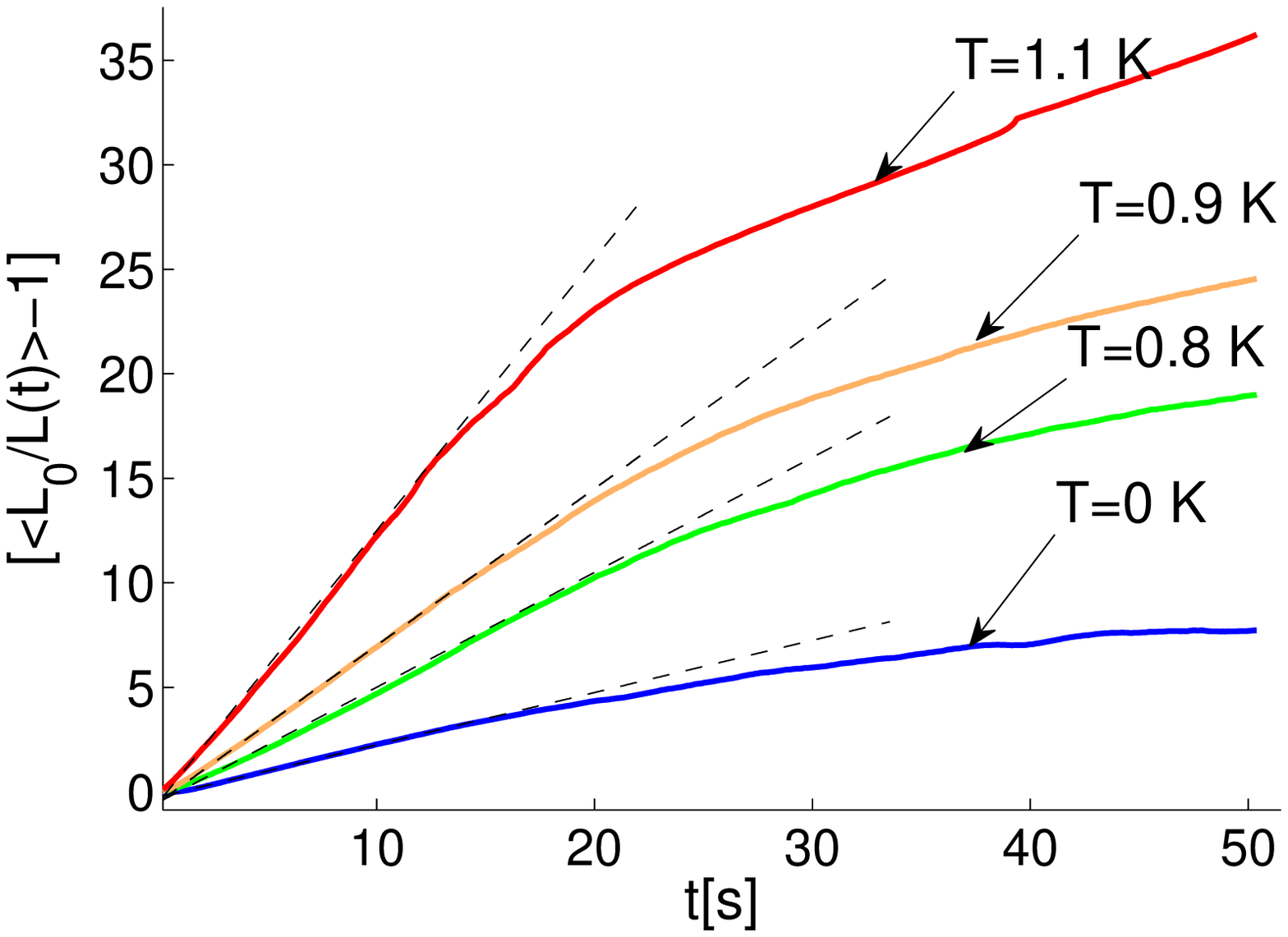} &
   \includegraphics[width=5.8 cm]{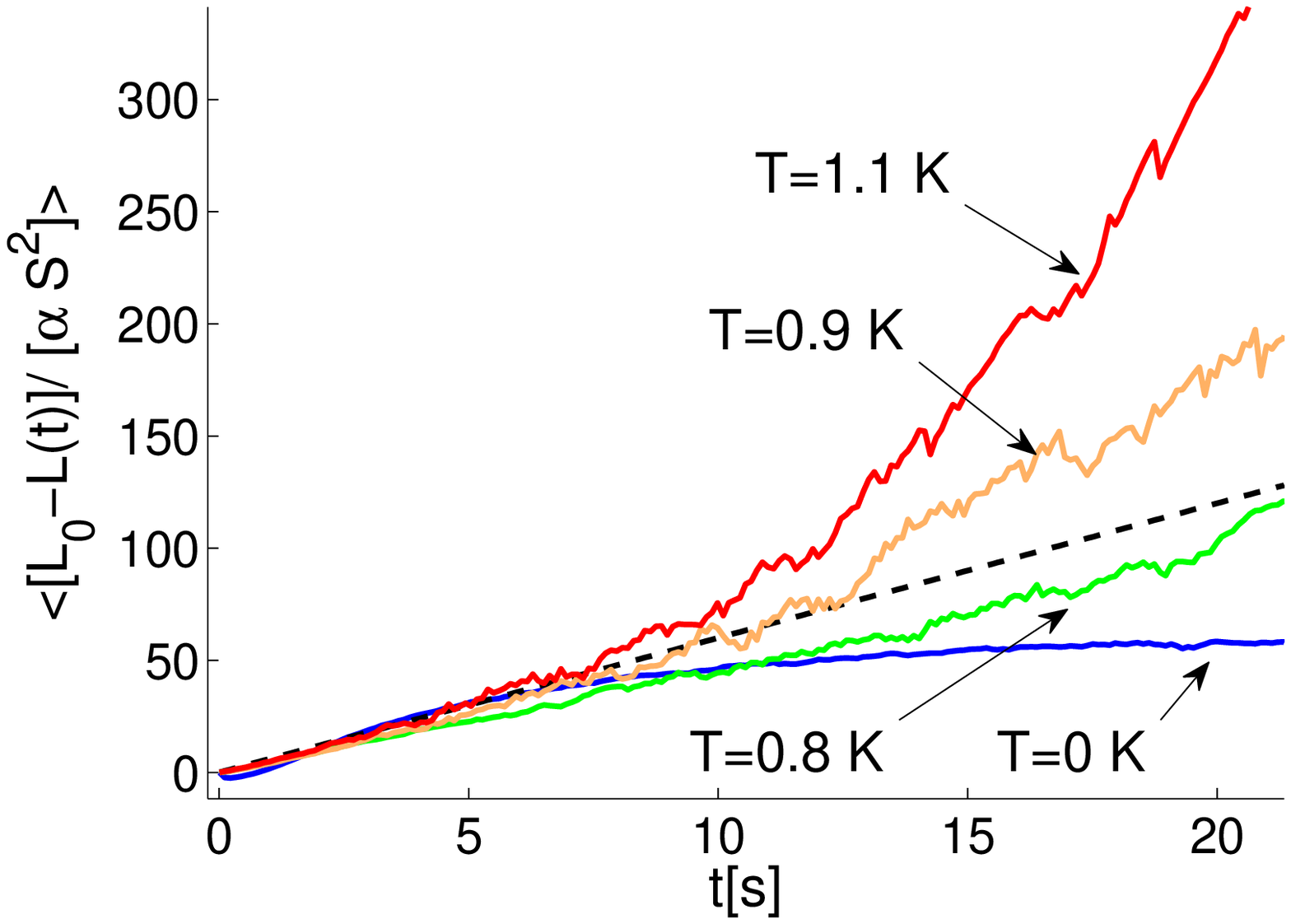}
   \\
  \hline
\end{tabular}
\caption{\label{f:2}Color online. \emph{Panel A.} The time dependence of normalized VLD  $\< \C L(t)/\C L_0\>$, averaged over 32 initial configurations,  for $T=0\,$K (blue line), $T=0.8\,$K (green line), $ T=0.9\,$K (brown line) and for $T=1.1\,$K (red line). The horizontal  dash-dotted red line shows the background VLD $\C L\sb {bg}=100\,$cm$^{-2}$. The black dashed line shows  the asymptotical decay  law $\C L(t)\propto t^{-1}$. \emph{Panel B.} The evolution of the VLD in the rectifying coordinates\,\eqref{V2}, $[\< \C L_0 /\C L(t)\>-1] $ versus $t$.  \emph{Panel~C.}~The evolution of the VLD in another  rectifying coordinates\, $\< \C L_0 /\C L(t)\> $ versus $t$, dictated by Eqs.\,\eqref{V2} and \eqref{chi2}: $[\C L_0 - \C L(t)]/ [\alpha  \widetilde  S^2 ]$
}
\end{figure*}

\section{\label{s:results} Temporal decay of the vortex tangle}
The temporal decay of the vortex tangle at $T=1.9\,$K was studied in Ref.\cite{27}. In this paper we show that   new  physics emerge  at much lower temperatures, $T\leq 1.1\,$K, including the zero-temperature limit. At these temperatures
 the entire evolution of the vortex tangle from the initial VLD
$\C L_0\simeq 6000\,$cm$^{-2}$ to the background level  $\C L\sb{bg}\simeq 100\,$cm$^{-2}$ takes between 50 to 100\,s.

To analyze the physics of the decay of the vortex tangle it is customary to begin with the traditional VLD dynamics. To provide fuller information we consider in this Section also the time dependence of the rate of reconnection
events, and of the curvature of the vortex lines. It turns out that the last quantity is most relevant in the
context of the Kelvin waves.

\begin{figure*}
\begin{tabular}{|c|c|c|}
  \hline
A & B & C\\
 \includegraphics[width=5.8 cm]{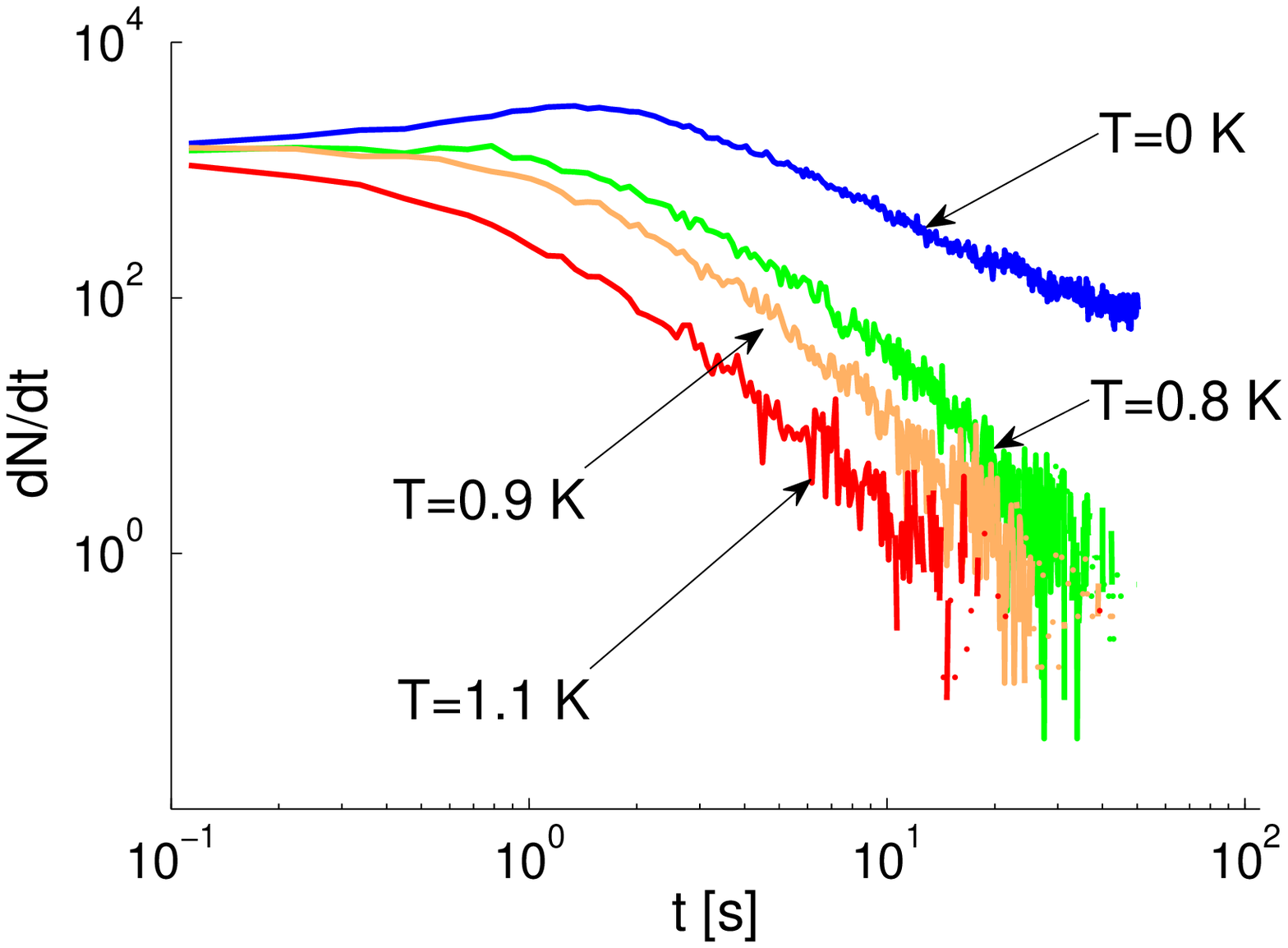}   &
 \includegraphics[width=5.8 cm]{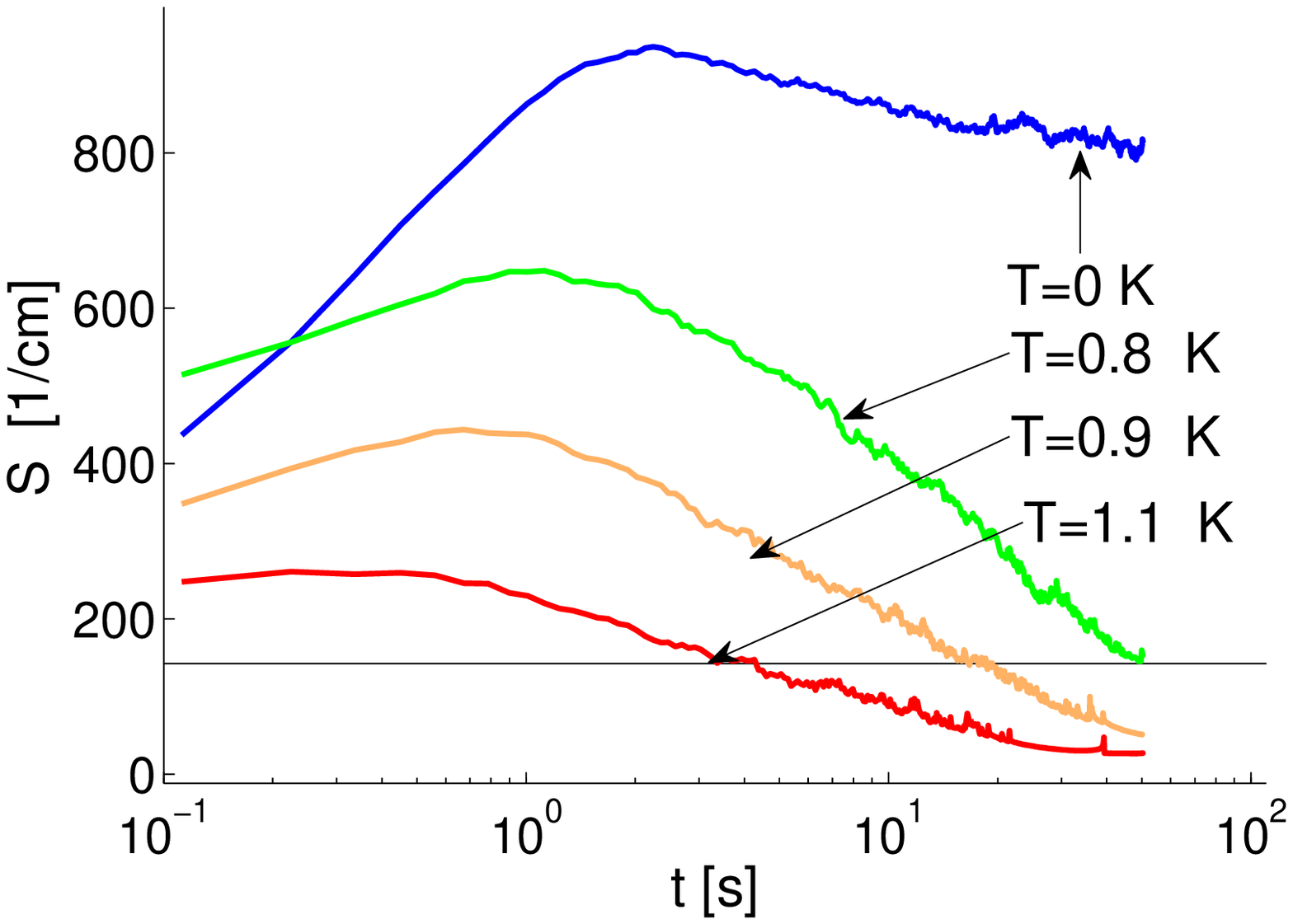} &
 \includegraphics[width=5.8 cm]{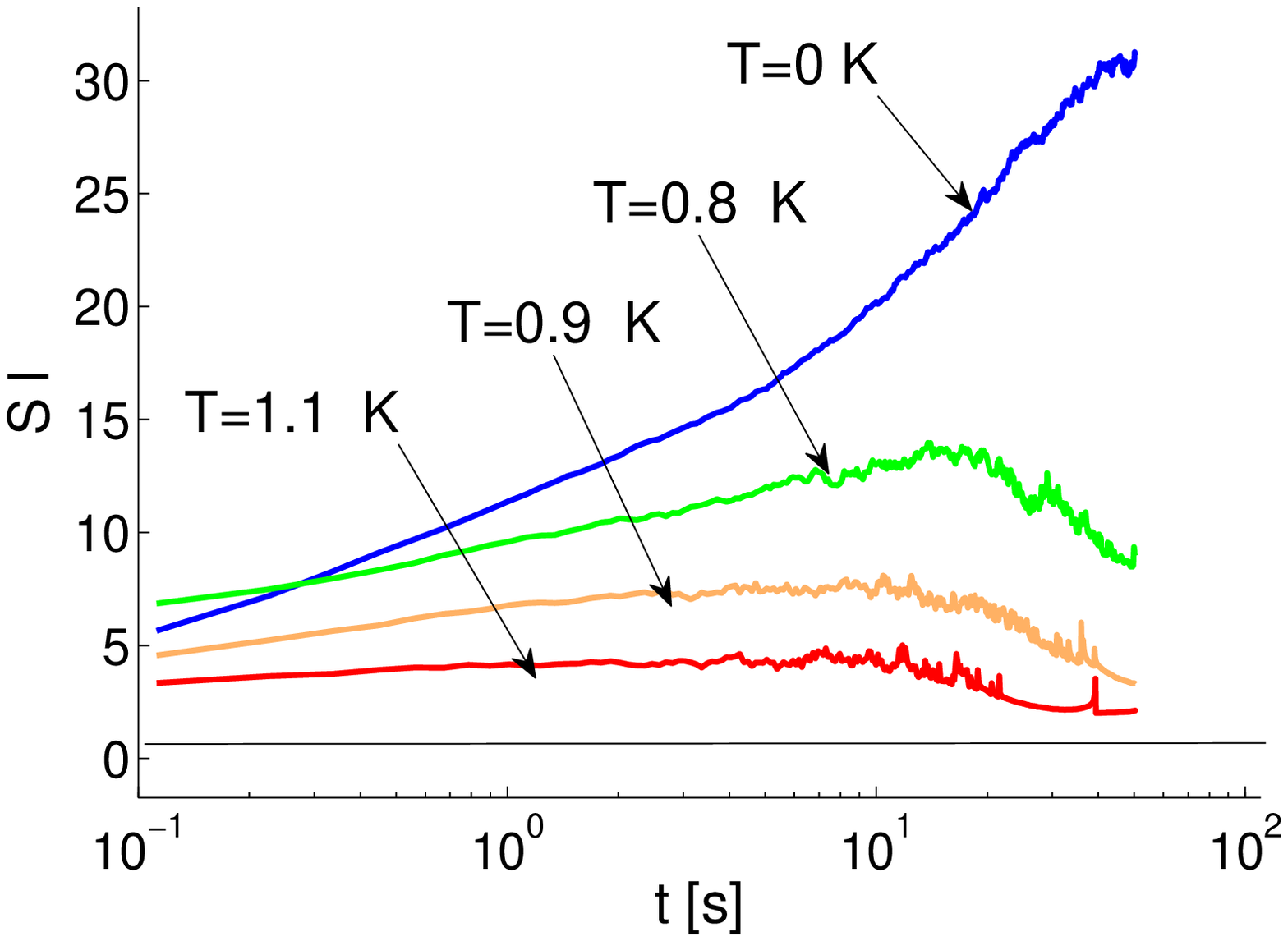}\\
  \hline
\end{tabular}
\caption{\label{f:3}Color online. \emph{Panel A.} The time dependence of the mean reconnection rate $\<d N(t)/ dt\>$ (in s$^{-1}$) for the same temperatures as in Figs.\,\ref{f:1}. \ref{f:2}.
\emph{Panel B.} The evolution  of the rms vortex line curvature  $ S(t)\equiv \sqrt {\< |s^{\prime\prime}|^2\> }$  at different temperatures. \emph{Panel~C.}
 The evolution of the dimensionless curvature $\ell(t)S(t)$,  normalized by the current intervortex distance  $\ell(t)$.   Thin horizontal lines show  initial curvature (dimensional in Panel B and dimensionless in Panel C).}
\end{figure*}

\subsection{\label{ss:VLD} Evolution of the vortex line density  and the effective viscosity}
The decay of the VLD is usually discussed in the framework of the Vinen equation. Without counterflow velocity\,\cite{Vin57}, it reads as:
\begin{equation}\label{V1}
\frac{d \C L(t)}{d t}= - \frac{\chi_2 \kappa \,\C L^2}{2\pi} \ ,
\end{equation}
where $\chi_2$ is a phenomenological coefficient.
The solution to this   equation is
\begin{equation}
\frac{ \C L_0}{ \C L(t)} -1= \frac{\chi_2  }{2\pi} \, \kappa   \C L_0 \, t\equiv b \, t\ .
\label{V2}
\end{equation}
In Fig.\,\ref{f:2} we present  the ensemble averaged results for the evolution of the VLD $\C L(t)$ for different temperatures and in different coordinates. Figure\,\ref{f:2}A shows plots of $\log \big[\< \C L(t) /  \C L_0\>\big ]$ with the black dashed line $\C L \propto 1/t$, corresponding to the large $t$ asymptotic of the solution\,\eqref{V2}.  Figure\,\ref{f:2}B shows the same evolution of VLDs, but in rectifying coordinates that are suggested by Eq.\,\eqref{V2}.
 One sees that at high temperatures the decay of the VLD proceeds faster  and reaches lower level. For times $t  \lesssim 15\, $s,  these results prove to be consistent with the predictions\,\eqref{V2} of the Vinen Eq.\,\eqref{V1} for the unbounded vortex tangle
   with the  slope $b  \approx 1.2,\,  0.75, \ 0.5$ and $0.25\,$s$^{-1}$ for $T=1.1,\ 0.9,\ 0.8$ and $0\,$K respectively.
  For longer times,  when the VLD decay such that   the intervortex  distance $\ell$ became compatible with the box size $H$, this assumption is violated  and the numerical results deviate from the prediction\,\eqref{V2}.   It is interesting to note that our  numerical results for $ T \leqslant 1.1\,$K, shown  in Fig.\,\ref{f:2}B, deviate down from the corresponding straight lines, while for the higher temperature $T=1.9\,$K studied in Ref.\,\cite{VTM03},   they deviate upward.

To clarify the temperature dependence of the decay consider the phenomenological coefficient $\chi_2$ which  is related\,\cite{Schwarz88,Lad03}  to the mutual friction parameter $\alpha$ and the rms vortex curvature $\widetilde S$  \footnote{The notation for this object, $\widetilde  S$,  is the same as in our Ref.\,\cite{KLPP14} to distingue it from the mean curvature, $\overline S\= \< |s''|\>$. }  :
\begin{eqnarray} \label{chi2}
\chi_2&=& \frac{\alpha  \Lambda}  2   \, (\ell \widetilde S)^2\,,\\
\label{curvA}
  \widetilde  S^2 &\equiv&  \frac 1 L \int_{\C C} |\bm s''|^2\, d\xi\ .
\end{eqnarray}
Here  $\bm s''(\xi)=  d^2 \bm  s/  d \xi^2$ is  the  local
curvature vector of the vortex line.
The time evolution of  $\widetilde S(t)$ at different temperatures will be discussed below in Sec.\,\ref{ss:curv}.

To check prediction of Eqs.\,\eqref{V2} and \eqref{chi2} we rewrote them as follows:
\begin{equation}\label{V3}
 \frac{ \C L_0 -\C L(t) }{ \alpha \widetilde S^2 } = \frac{\kappa \C L_0}{2\pi} \, t\,,
\end{equation}
and plotted in Fig.\,\ref{f:2}C its left-hand side (after ensemble averaging) versus $t$. The black dashed line has  the temperature independent slope  $\kappa \C L_0/2\pi$ in Eq.\,\eqref{V3}. One sees that indeed for $t  \lesssim 7\,$s all the lines collapse as expected from Eq.\,\eqref{V3}.  Notice that for zero temperature $\alpha=0$, while for the $T=0\,$K plot (the solid blue line) we choose $\alpha=\alpha\sb{eff}\simeq 1.2 \cdot 10^{-4}$, which corresponds to $T\simeq 0.68\,$K  according to Eq.\,\eqref{alpha}.

Notably, we observe at $T=0\,$K a decay which
is similar to the one observed at higher temperatures. Thus the decay cannot be caused by mutual friction. It is a common belief   that the main candidates for the energy dissipation at $T\to 0\,$K are the  direct  Kelvin wave energy cascade and the loop self-crossing that create smaller and smaller loops\,\cite{S95,TAN-2000,V01,KBL11}.
Indeed, at the later stage of the tangle evolution, shown  in Fig.\,\ref{f:1} at $t=50\,$s, there are  only smooth vortex lines for $T\geq 0.8\,$K (Figs.\,\ref{f:1}B-\ref{f:1}D), while the vortex lines for $T=0\,$K (Panel E) are very wrinkled and there are some small loops. A fragment of this configuration, shown in Fig.\,\ref{f:1}F for three successive times clearly demonstrate the propagation of short Kelvin waves along the vortex line (visible as its meandering around a smooth line) and the fast motion of almost invariant small loops.

 Numerical results shown in Fig.\,\ref{f:2}B allow us to estimate the effective (Vinen's) kinematic viscosity $\nu'(T)$ defined via the rate of energy dissipation $\varepsilon$:
\begin{equation}\label{nup}
\varepsilon = \nu'(T) (\kappa \C L)^2\simeq \nu'(T)\, \frac {\kappa^2}{\ell^4}\ .
\end{equation}
According to Walmsley-Golov Ref.\,\cite{WG} the free decay of the VLD depends on $\nu'$ as follows
\begin{equation}\label{Lt}
\C L(t) = \frac{\Lambda}{4\pi}\,\frac 1{\nu'\, t}\ .
\end{equation}
Comparing Eqs.\,\eqref{V2} and \eqref{Lt} one concludes that the numerical values of $b$ and $\nu'$ should be related:
\begin{equation}\label{rel1}
b= 4 \pi \,\nu' \C L_0/\Lambda\ .
\end{equation}
 Taking our numerical values $\C L_0\approx 6\cdot 10^3\,$cm$^{-2}$ and $b\approx 1.2\,$s$^{-1}$,\ $0.75\, $s$^{-1}$ and $b\approx 0.5\,$s$^{-1}$ for $T=1.2\,$K, \ $0.9\,$K and $0.8\,$K, respectively, together with  $\Lambda /(4\pi)\approx 1.2$,  as in Ref.\cite{WG}, we estimate the ratio $\nu'/\kappa\approx 0.24\,,\ 0.15$ and $0.$1 for these temperatures. This estimate is in close agreement with the experimental observations, see Fig.\,5 in Ref.\,\cite{WG}. The comparison in the zero-temperature limit requires  careful analysis of the effect of finite numerical resolution and will be done elsewhere.

\begin{figure*}
\begin{tabular}{|c|c|c|}
  \hline
 A & B & C\\
 \includegraphics[width=5.8 cm]{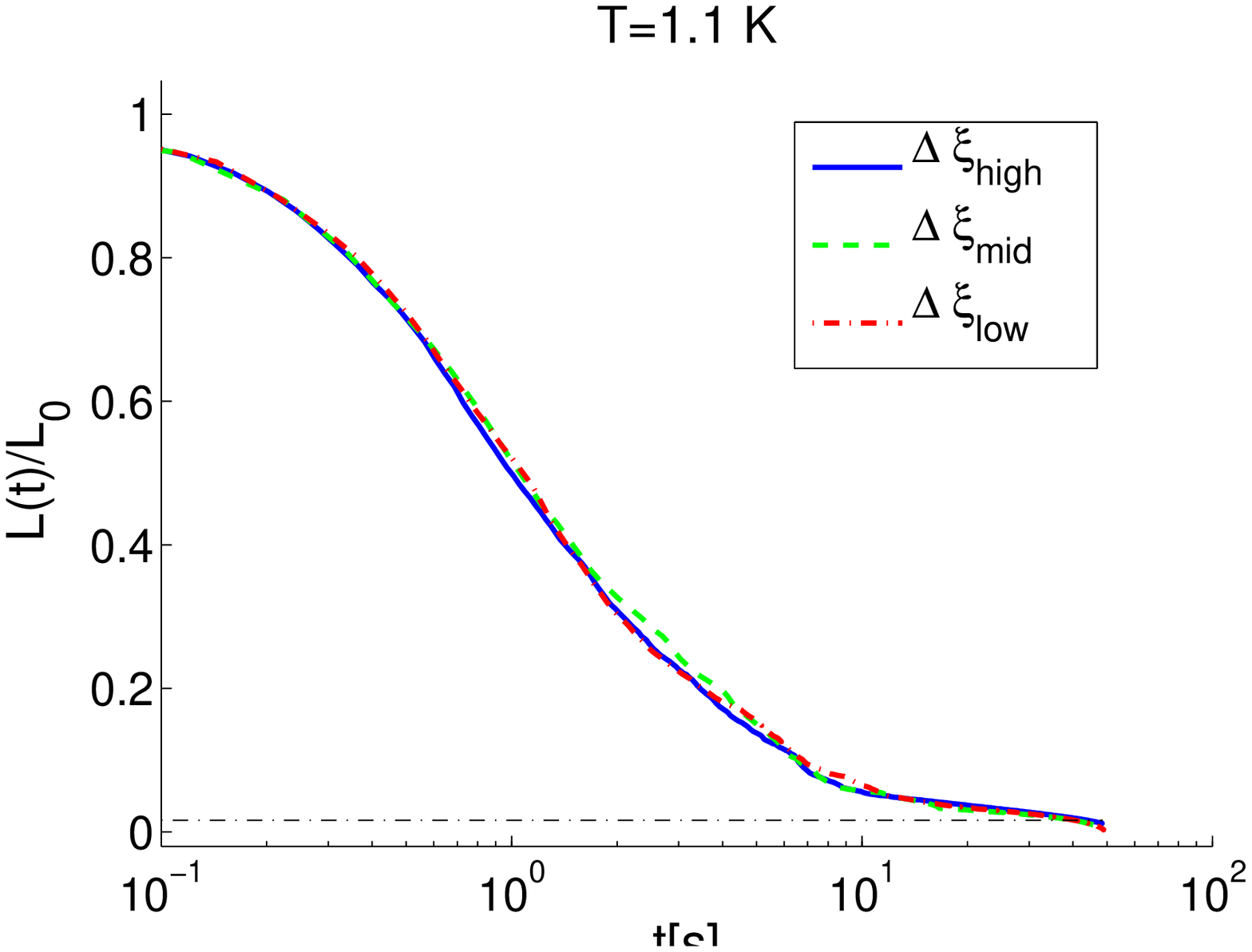}  &
   \includegraphics[width=5.8 cm]{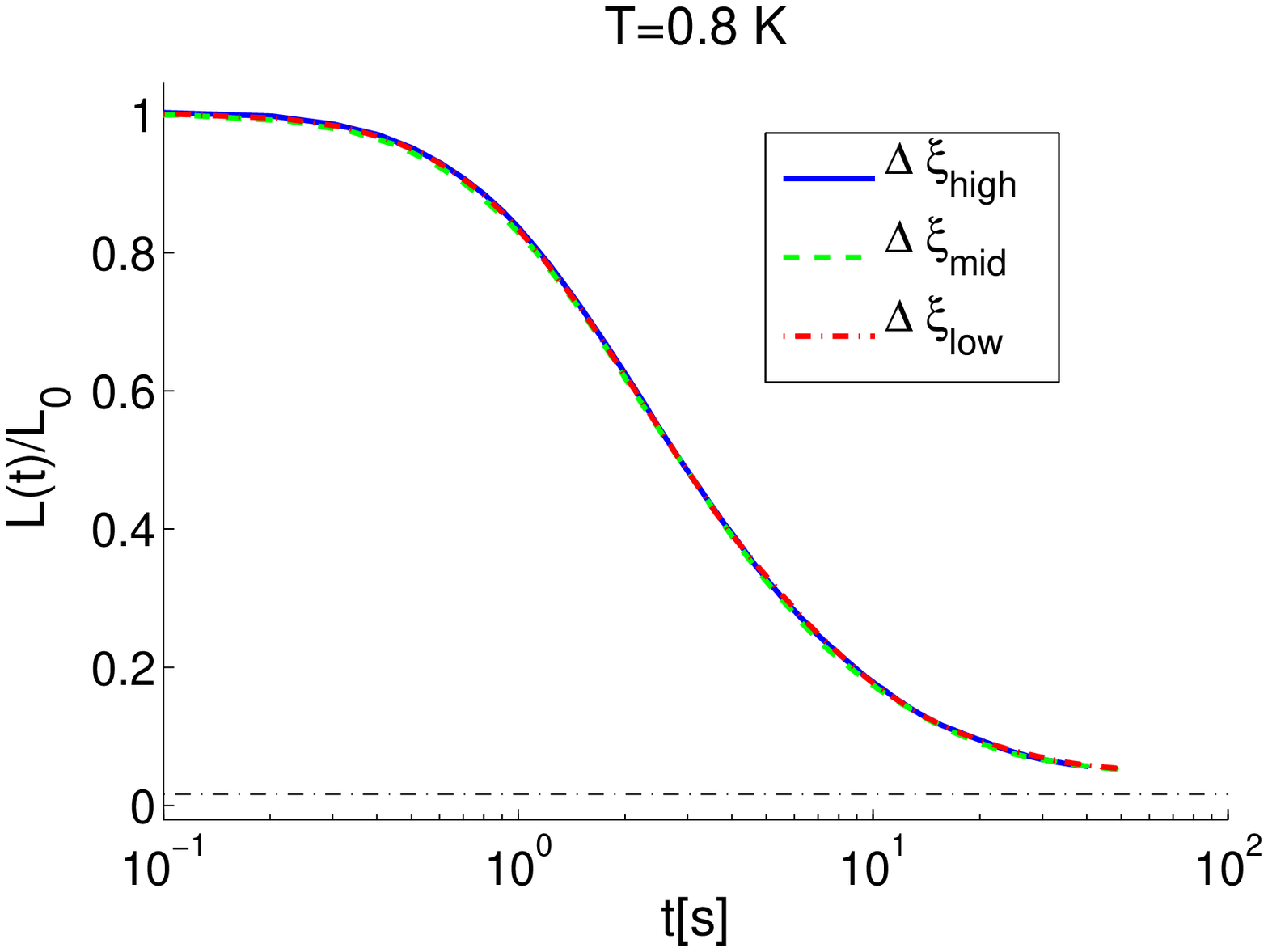} &
  \includegraphics[width=5.8 cm]{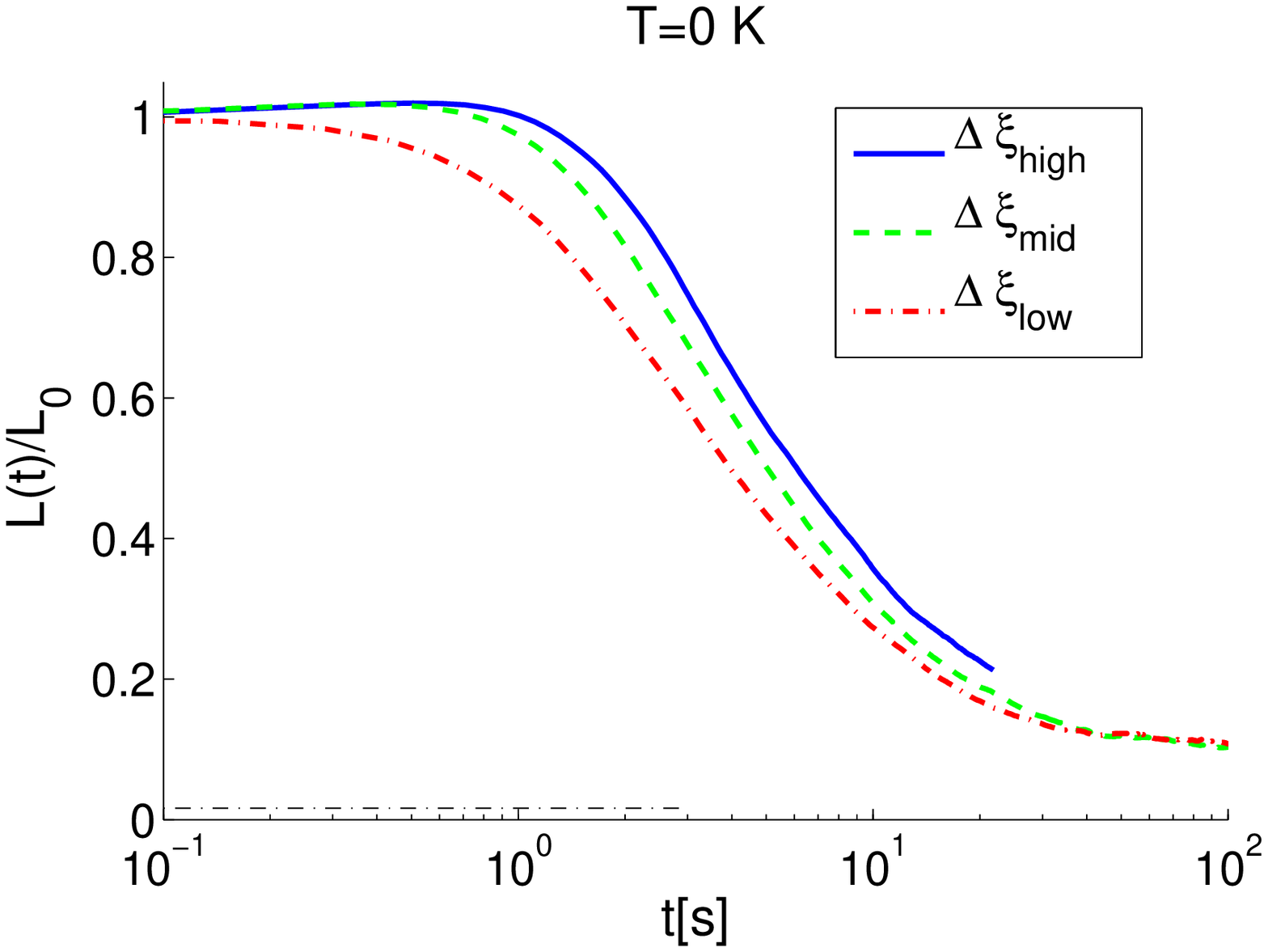}\\
  \hline
\end{tabular}
\caption{\label{f:4} Color online.  \emph{Panels A,B} and  \emph{C.}  Comparison of the time dependence of the normalized mean VLD  $\< \C L(t)/\C L_0\>$ for different resolutions and  for $T=1.1\,\ 0.8$K and $T=0$. The horizontal  dash-dotted red line shows the background VLD $\C L\sb {bg}=100\,$cm$^{-2}$.
}\end{figure*}

\subsection{\label{ss:recon} Evolution of the reconnection rate}

Reconnection events excite Kelvin waves on the vortex lines and in addition may create additional smaller loops by self-crossing.
Therefore an important characteristics of the evolution of the vortex tangle is the reconnection rate defined as the number of reconnection events per unit time, $dN(t)/dt$. The computational algorithm includes explicitly the reconnection procedure, thus allowing us to measure
this quantity. Figure \ref{f:3}A  shows that the reconnection rate at $T=0$ (blue line) is much larger than that at $T=0.8, 0.9\,$K and $T=1.1\, $K (green, brown and red lines). This is a combined consequence of two reasons: larger line density [see Fig.\,\ref{f:2}A] and faster motion of the vortex lines caused by their large curvature [see Fig.\,\ref{f:2}C]. The conclusion is that the excitation of
Kelvin waves and the production of small loops is ever more efficient when the temperature reduces down to zero.

\subsection{\label{ss:curv} rms of the vortex-line curvature and the energy spectrum of Kelvin waves}

 For our considerations the most important information about the level of Kelvin wave excitation is given by the  rms of the vortex line curvature $\widetilde  S$, defined by Eq.\,\eqref{V3}.
For small deviations of the vortex line $\delta x$ and $\delta y$  in the $\bm x$ and $\bm  y$ directions from the  straight line, oriented along the $\B z$-axis the vortex length can be approximated as
\begin{subequations}\label{cur}
\begin{eqnarray}\label{6B}
L\approx \int \sqrt {1+ |w'|^2} dz\,, \  w(z)\= x+i z\,, \  w'\= \frac{dw }{ dz} \, . ~~
\end{eqnarray}
 In the same approximation, $|w|\to 0$, $d \xi \approx d z$ and $|s''|\approx |w''|\= | d^2 w /d z^2|$. This allows one to approximate $\widetilde S$ for small amplitudes of Kelvin waves as follows:
\begin{equation}\label{6C}
(\widetilde S)^2\approx \frac 1 L \int |w''|^2  \, d z\ .
\end{equation}\end{subequations}
 In the local induction approximation the   energy density (per unite   mass) of the unit length of the vortex line   is $\kappa^2 \Lambda/(2\pi)$, where $\Lambda = \ln ({\ell}/a_0)$, (see e.g.\cite{S95}). Now, expanding the square root in Eq.\,\eqref{6C}, we obtain  the equation for  total energy density (per unit  length, per unit  mass) of the Kelvin waves:
  \begin{subequations}\label{curv1}
 \begin{equation}\label{7A}
 E \approx \frac  {\kappa^2 \Lambda}{4\pi\, L}   \int |w'|^2  \, d z\ .
 \end{equation}
This allows us to rewrite identically Eq.\,\eqref{6C} as follows:
\begin{equation}\label{KW-E}
  (\widetilde  S)^2  \approx  \frac{4\pi}{\kappa^2 \Lambda} E\, R\,, \quad R\= \frac {\int |w''|^2  \, d z}{\int |w'|^2  \, d z}\ .
 \end{equation}
The next step is to rewrite the integrals in the ratio $R$ in the $k$ representation
\begin{equation}\label{7C}
R= \frac{\int k^4 |w_k|^2 dk }{\int k^2 |w_k|^2 dk}\,,
\end{equation}
where $w_k$ is the Fourier image of $w(z)$. Bearing in mind that the canonical amplitudes of Kelvin waves, $a_k$, are proportional to $w_k$ and that the Kelvin wave spectrum $E(k)\propto k^2 |a_k|^2$ we present the ratio\,\eqref{7C} as
  \begin{equation}\label{7D}
R= \frac{\int k^2 E(k) dk }{\int E(k) dk}= \frac 1 E \int k^2 E(k) dk \ .
\end{equation}\end{subequations}
Substituting Eq.\,\eqref{7D} into the first of Eq.\,\eqref{KW-E}
we finally get the estimate of $(\widetilde S)^2$ in terms of the energy spectrum of the Kelvin waves:
\begin{equation} \label{curvB}
  (\widetilde  S)^2\simeq \frac{4\pi}{\Lambda \kappa^2} \int\limits_{k\sb{min}}^{k\sb{max}} k^2 E(k)\, dk\ .
\end{equation}
 Using the spectrum\,\eqref{LN}, which is presumably validin some interval of scales from the lower wave vector $k\sb{min}$ (which is about inverse box size) up to some large cutoff wave vector $k\sb{max}$ (about the inverse core size in the physical system or  the resolution scale in the simulations, see below) we get for the case of LN spectrum\,\eqref{LN} of   weak turbulence of Kelvin waves:
\begin{subequations}\label{Curv}
\begin{equation} \label{curvC}
  \ell \, \widetilde  S \simeq  \Phi\, (  \ell \, k\sb{max}  )^{2/3} \,,
   \end{equation}
   where
  \begin{equation} \label{curvC1}  \Phi \equiv \sqrt {4 \pi E  / ( \Lambda \,\kappa^2) }\,,
  \quad E=  \int\limits_{k\sb{min}}^{k\sb{max}} E(k)\, dk\ .
\end{equation}
 The same derivation with the KS-spectrum\,\eqref{KS} of  weak turbulence of Kelvin waves gives 
\begin{equation} \label{curvE}
  \ell \, \widetilde  S \simeq  \Phi\, (  \ell \, k\sb{max}  )^{4/5} \,,
\end{equation}
while in the case of strong turbulence with the Vinen's spectrum\,\eqref{V} we have:
   \begin{equation} \label{curvD}
  \ell \, \widetilde  S \simeq  \Phi\,   ( \ell \, k\sb{max} )  \ .
\end{equation}
\end{subequations}
Note  that in any case the normalized curvature $ \ell \, \widetilde S $ is a small-scale effect and no saturation of  $ \ell \, \widetilde S $ is expected with increasing  the line resolution. Notice also
that two parameters determine $(\widetilde S)^2$: the total energy of Kelvin wave excitations, entering the Eq.\,\eqref{curvC} via the dimensionless parameter $\Phi$, and the dimensionless upper cutoff $\ell k\sb{max}$ of the power-like Kelvin wave energy spectrum. Indeed, without Kelvin waves propagating along the straight vortex lines the curvature should vanish, i.e.  $(\widetilde S)^2=0$. In addition, it is reasonable that the curvature of the vortex line should be dominated  by the shortest Kelvin wave in the system, i.e. by $k\sb{max}$, the parameter that appears in the estimate\,\eqref{curvC} due to the divergence of the integral.

\begin{figure*}
\begin{tabular}{|c|c|c|}
  \hline
  A & B & C\\
 \includegraphics[width=5.8 cm]{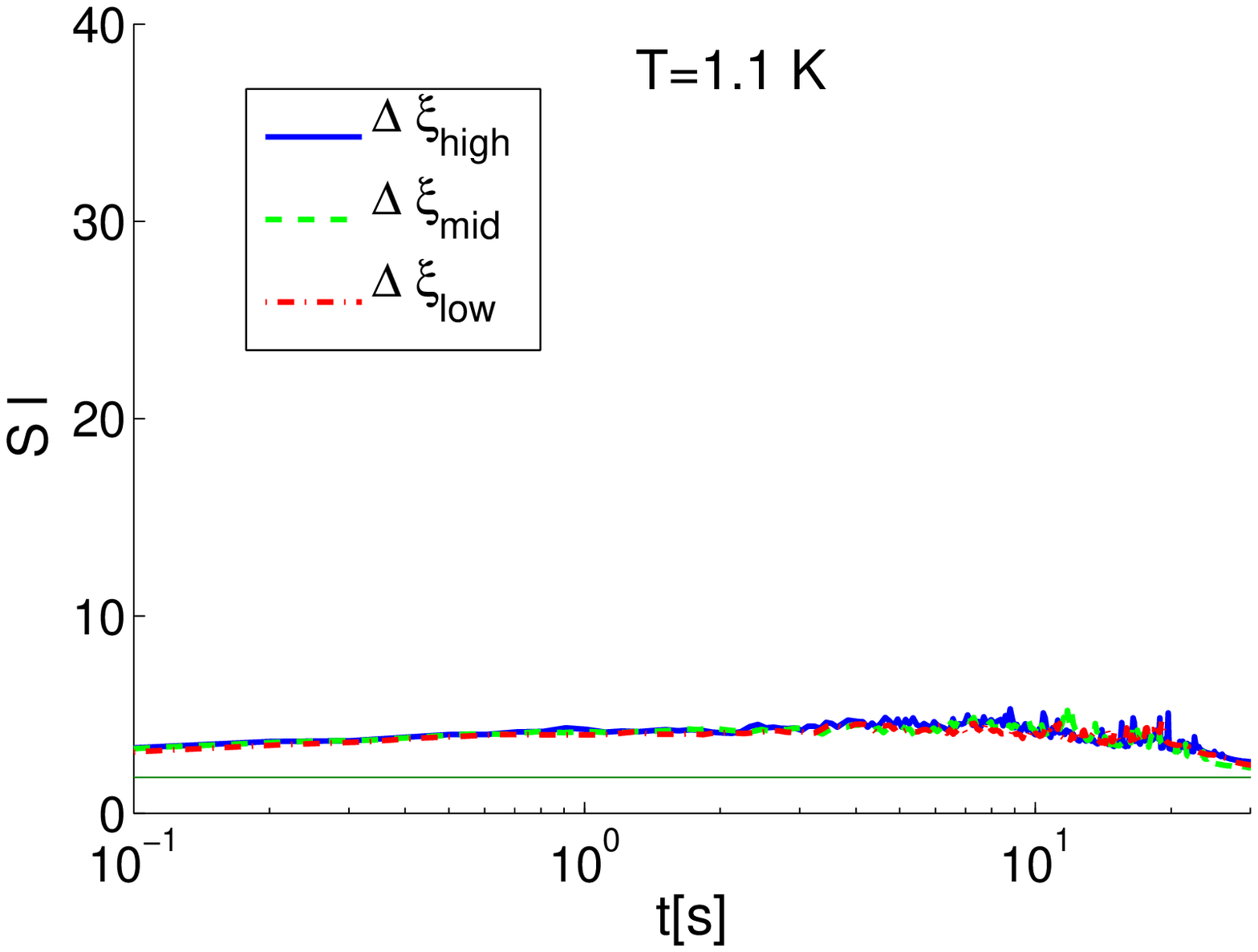}  &
 \includegraphics[width=5.8  cm]{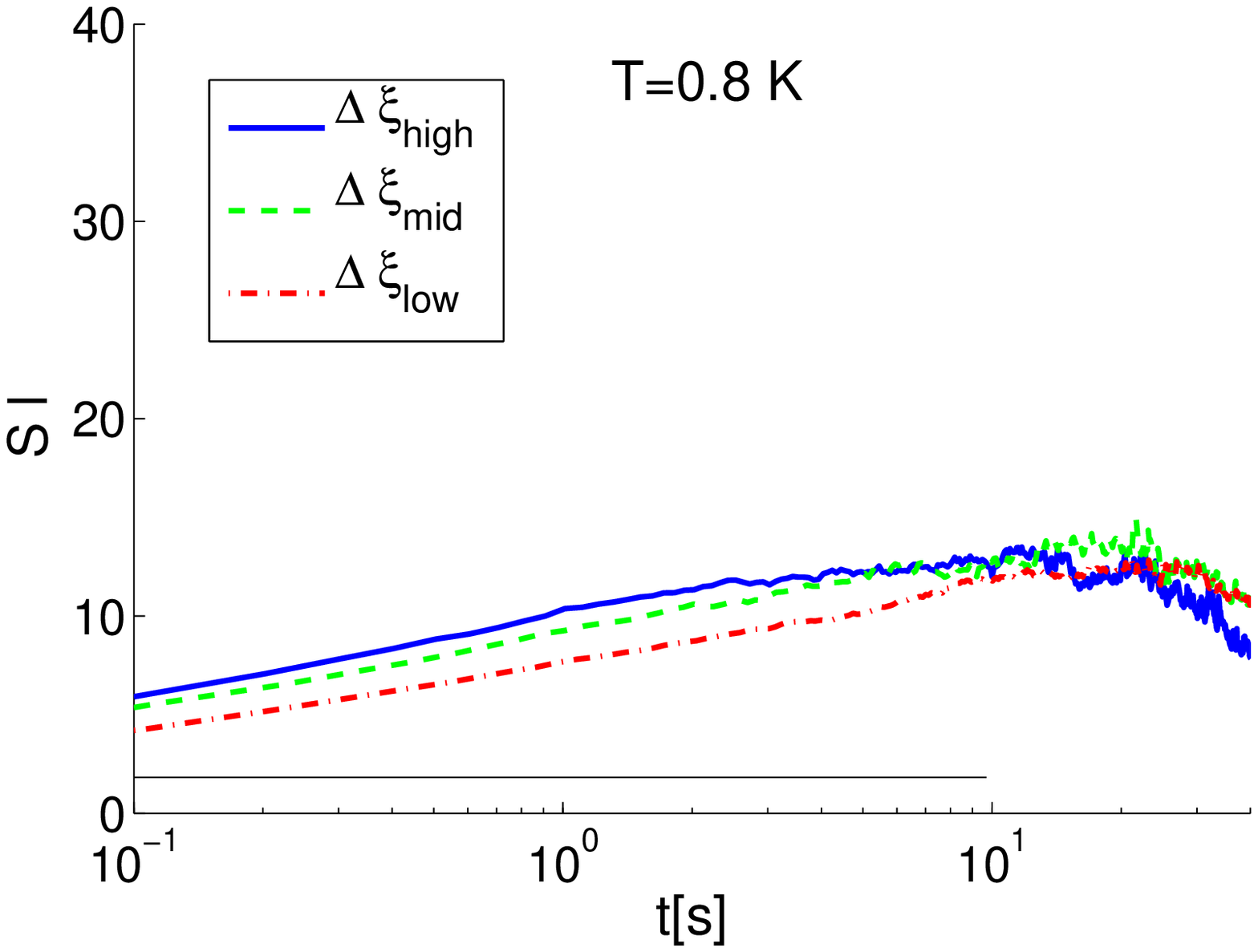} &
 \includegraphics[width=5.8  cm]{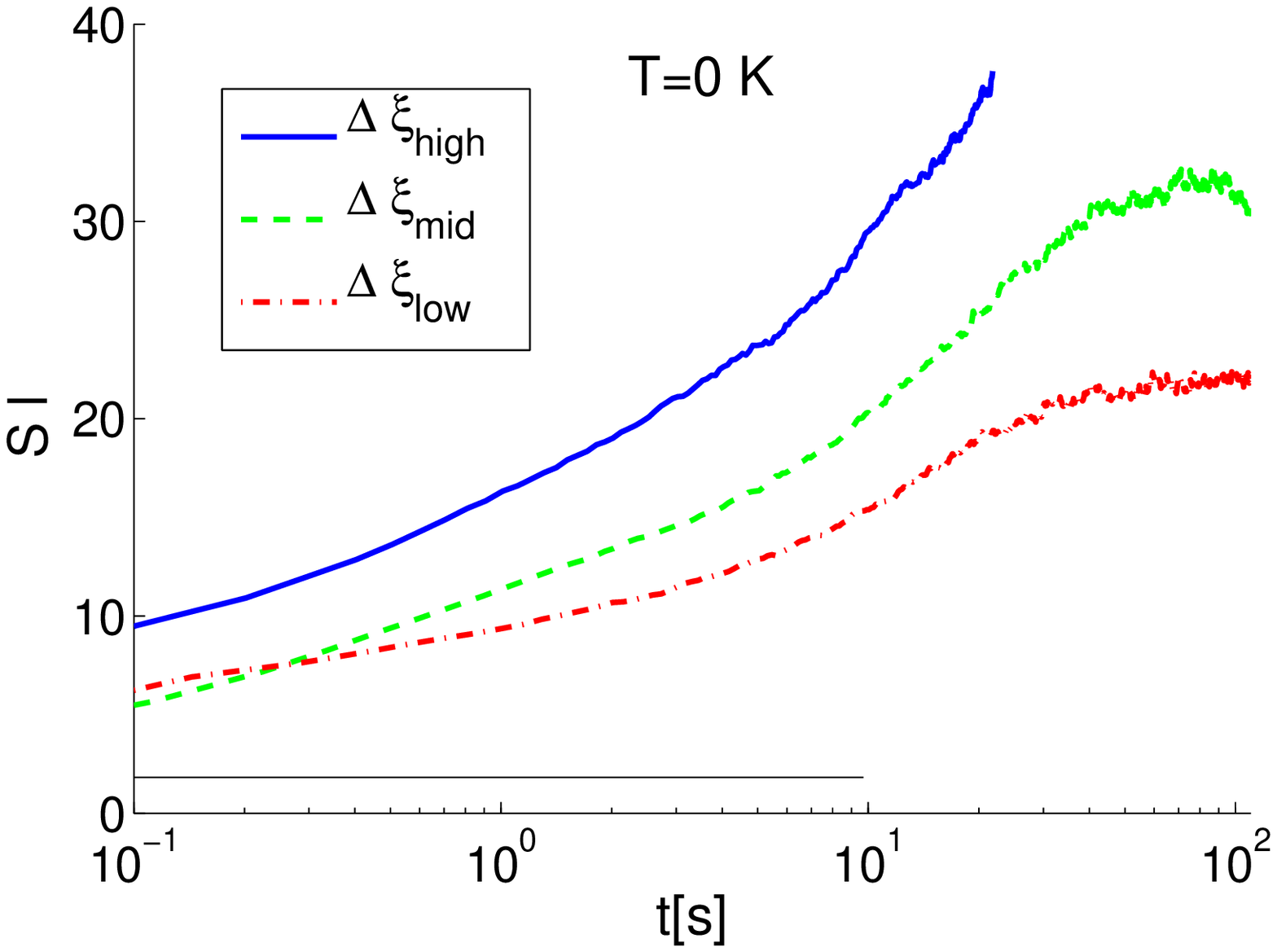}\\
  \hline

\end{tabular}
\caption{\label{f:5}Color online.
 Comparison of the time dependence of the normalized curvature  $ \ell(t) \widetilde S$  for different resolutions and the same temperatures.
}\end{figure*}
\begin{figure*}[t]
\begin{tabular}{|c|c|c|}
  \hline
 A & B & C \\
 \includegraphics[width=5.7 cm]{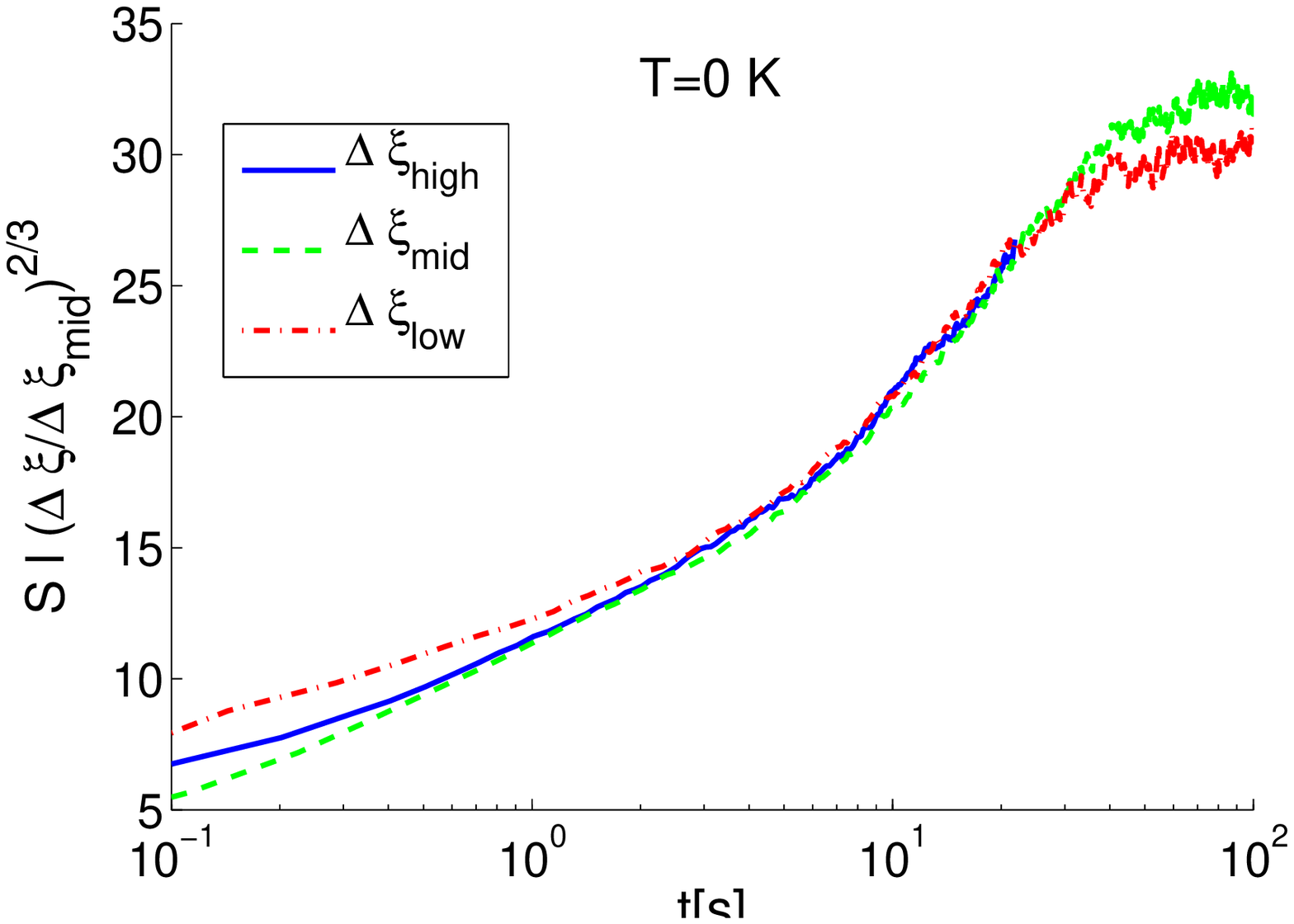}&
\includegraphics[width=5.7 cm]{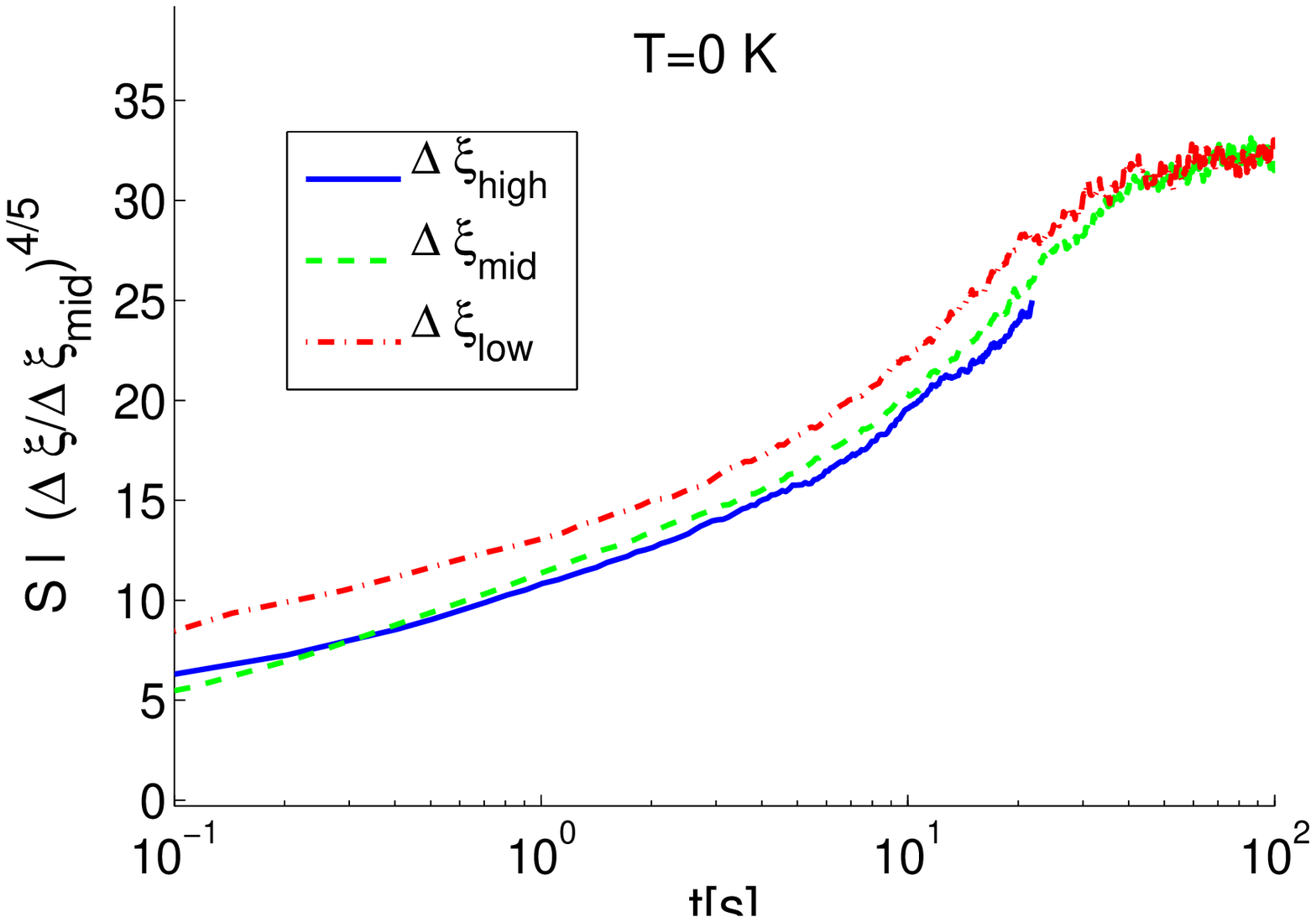}&
 \includegraphics[width=5.7 cm]{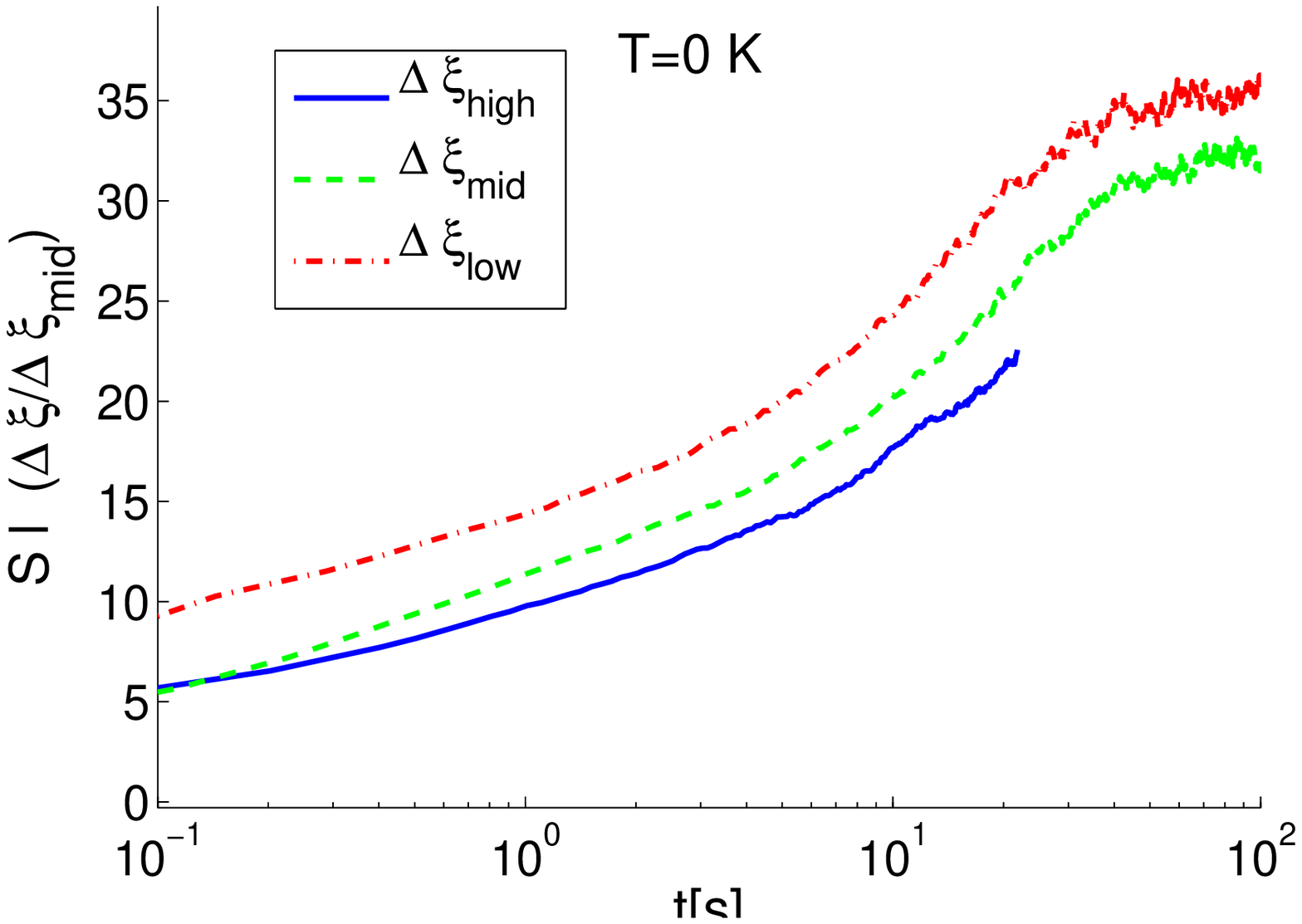}\\ \hline
  \end{tabular}
\caption{\label{f:6}Color online.  Evolution of the vortex line curvatures $\ell \widetilde S $ at different scale resolutions  compensated by $(\Delta \xi/ \Delta \xi\sb {mid})^ {2/3}$ (\emph{Panel A)},  by $(\Delta \xi/ \Delta \xi\sb {mid})^ {4/5}$ (\emph{Panel B)}, and by  $(\Delta \xi/ \Delta \xi\sb {mid})$ \emph{(Panel C}).}
\end{figure*}
\subsection{\label{ss:curvA} Evolution of the mean vortex-line curvature}
 Armed with the estimate\,\eqref{curvC},  we  can consider now the numerical results for $\widetilde S(t)$ at different temperatures, shown in Fig.\,\ref{f:3}B, and for $\ell (t)\widetilde S(t)$ -- in Fig.\,\ref{f:3}C. In these plots,  the initial values   $\widetilde S(0)$ and $\ell (0)\widetilde S(0)$, taken from the stationary vortex configuration at $T=1.9\,$K,  are shown by a horizontal solid thin black line. During the first second, when the VLD is practically constant,  $\C L(t)\simeq \C L(0)$ (see Fig.\,\ref{f:2}A), the curvature for $T= 1.1\,$K roughly doubles (red line in  Fig.\,\ref{f:2}C) and roughly triples for $T=0.9\,$K. The effect is even larger for $T=0.8\,$K. The largest increase of  $\widetilde S(t)$ corresponds to zero temperature, shown by the blue line in  Fig.\,\ref{f:2}C.  We relate this effect to the increasing level of excitations of Kelvin waves at lower temperatures, partially due to the increase of the reconnection rate, as shown in Fig.\,\ref{f:2}B,  and partially due to the decrease of Kelvin wave damping due to the mutual friction (see Table \ref{t:1}).

During further evolution of the vortex tangles for $t> 1\,$s  the curvature $\widetilde S(t)$ of the tangles at finite temperatures $T> 0\,$K  decays, demonstrating a decrease of the  Kelvin waves energy,  caused by the decrease of the reconnection rate (see Fig.\,\ref{f:2}B) which is responsible for the Kelvin waves excitations. Nevertheless, the product $\ell (t)\widetilde S(t)$, as  Fig.\,\ref{f:2}D shows,  continues to grow during the first $10\,$s because of the fast growth of $\ell = 1/\sqrt {\C L}$. Finally, at later times, $t\sim 10\,$s, the curvature $\widetilde S(t)$ for $T> 0\,$K  decays even below its initial value and $\ell (t)\widetilde S(t)$ decays to about its initial value $\ell (0)\widetilde S(0)$. Accordingly, the configurations Figs.\,\ref{f:1}B--D consist of very smooth vortex lines. The vortex tangles show a completely different behavior at $T=0$, when there is no mutual friction: the tangles decay is much slower (Fig.\,\ref{f:1}A), the reconnection rate is much larger  (Fig.\,\ref{f:1}B). The main difference between zero-temperature and $T>0\,$K cases are the large curvature of the vortex lines that saturates at level $\ell \widetilde S \simeq 30$, exceeding substantially the initial level $\ell_0 \widetilde S_0 \simeq 2$.

The results for $T=0.43\,$ and $0.55\,$K (not shown in  the figures) are very close to those for $T=0\,$K.

\subsection{\label{ss:est} Estimating the upper cutoff of the Kelvin wave spectrum}

With the numerically found value $\ell \widetilde S\simeq 30$ for $T=0\,$K, we can estimate the upper cutoff $k\sb{max}$ of the Kelvin wave spectrum, using Eq\,\eqref{curvC}. Without large scale motion the total energy density (per unit length) of the tangle is given as $E=\kappa^2 \Lambda/4\pi$ (see e.g. Ref.\,\cite{72}). The data shown in Fig.\,\ref{f:1} allows a rough estimate of the length of
 straightened vortex lines (i.e. without Kelvin waves) as a half of the total (curved, i.e. with Kelvin  waves) vortex lines. This gives $E\simeq E/2\simeq \Lambda \kappa^2/(4\pi)$. Thus $\Phi\simeq 1$ in Eq.\,\eqref{curvC} and consequently $\ell\,  k\sb{max}\simeq (\ell \widetilde S)^{3/2}\simeq 160 \,$cm$^{-1}$  [for the LN spectrum\,\eqref{LN}].

Obviously, the shortest wavelength $\lambda_{min}$ resolvable in our simulation is about $2\Delta \xi$. Thus the highest resolvable
wavevector is $k\sb{res}\simeq \pi/ \Delta\xi\simeq 260\,\,$cm$^{-1}$. The consequence is that the spectrum of Kelvin waves cascaded
practically all the way down to the numerical resolution. This understanding calls for a careful check of the dependence
of the curvature at $T=0$ on the spatial resolution of the simulation. This is dealt with in the next subsection, resulting
in a demonstration that (i) the Kelvin waves are statistically important and (ii) the Kelvin wave dynamics is indeed weakly turbulent.

\begin{figure*}[t]
\begin{tabular}{|c|c|}
  \hline
 A & B \\
\includegraphics[width=5.8cm]{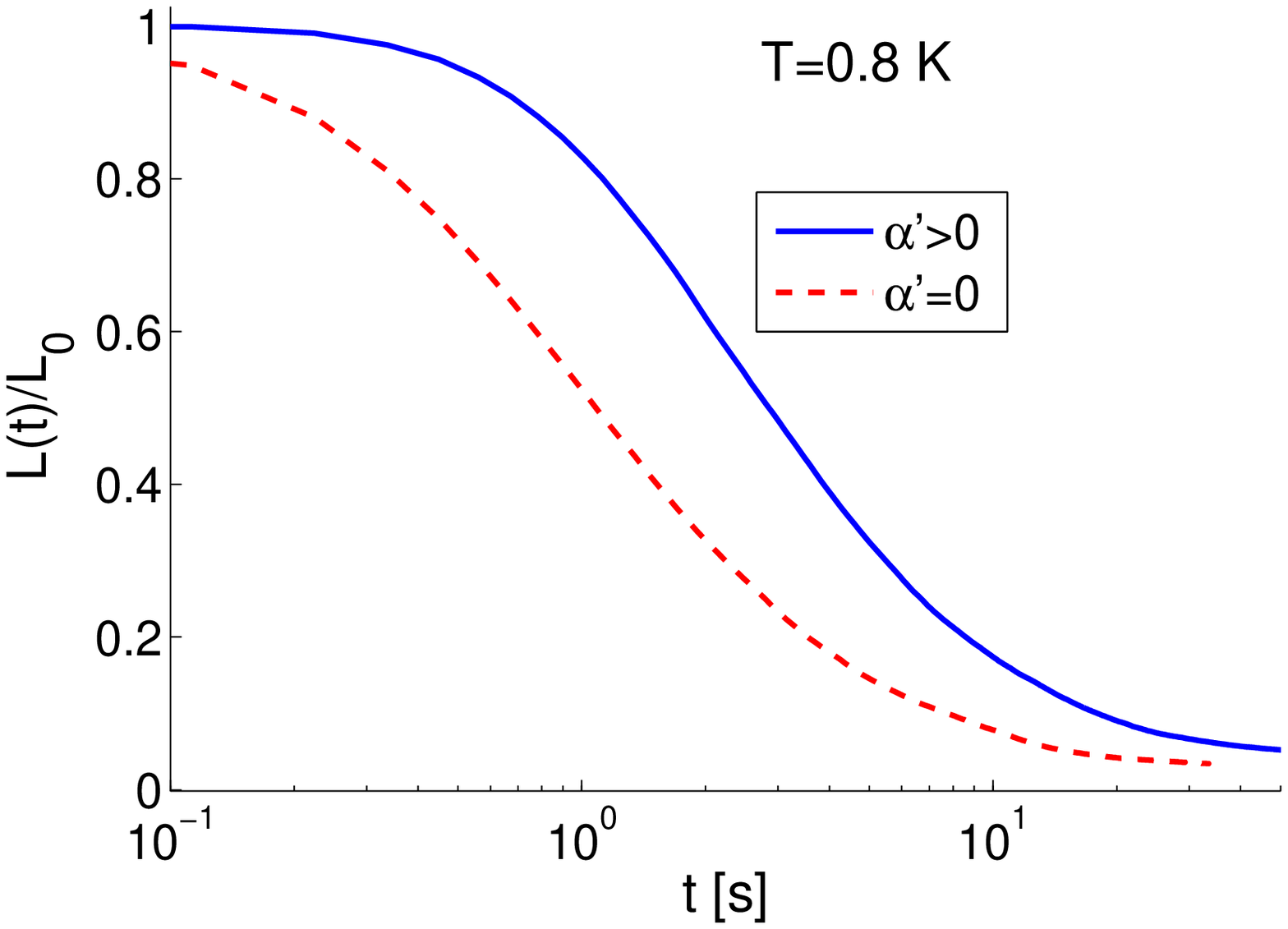}&
 \includegraphics[width=5.8 cm]{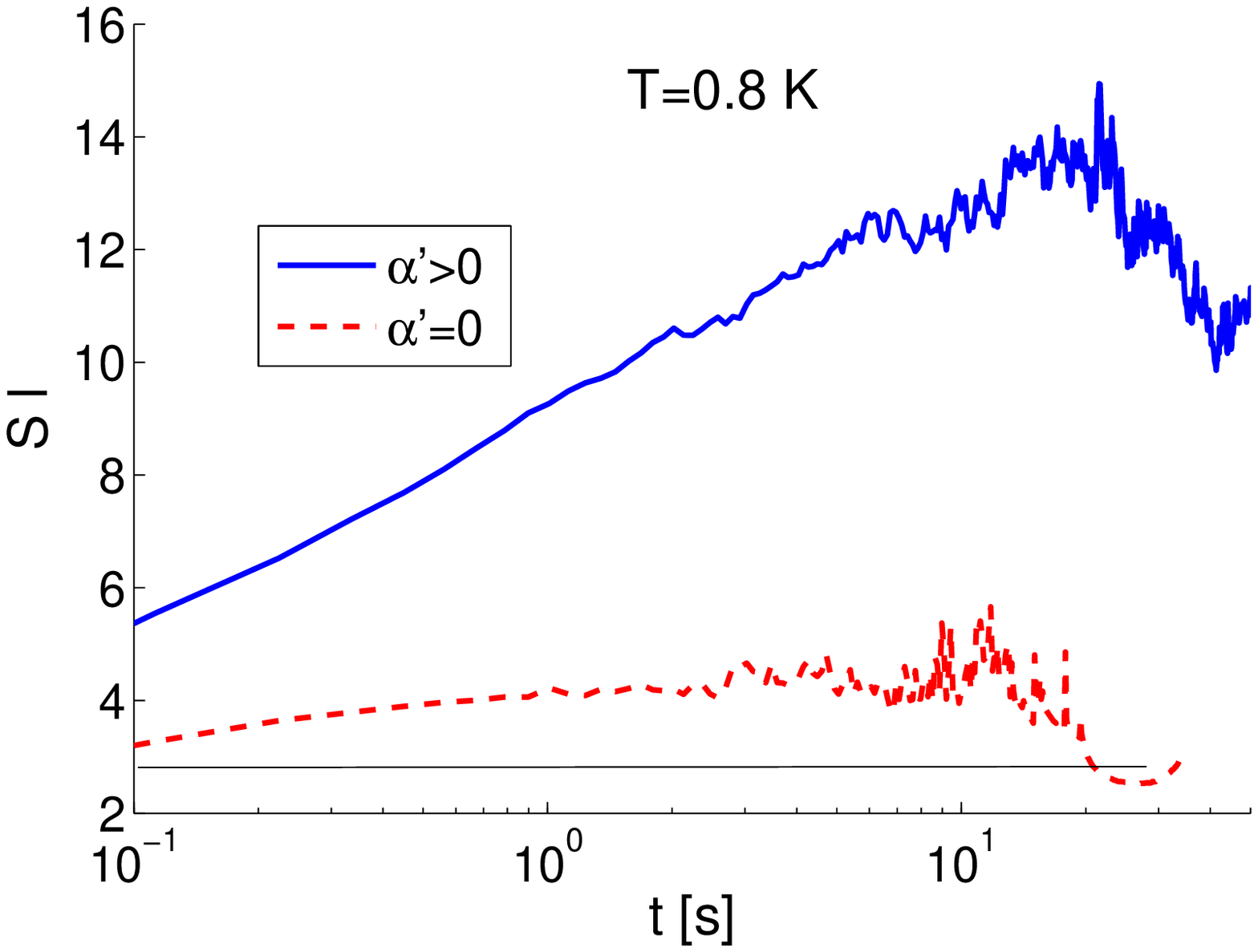}\\ \hline

 \end{tabular}
\caption{\label{f:7}Color online.  Comparison of the evolution of the normalized VLD, $\C L(t)/\C L(0)$ (\emph{Panel A}) and of the normalized curvature $S(t)\ell(t)$ (\emph{Panel B})  for the simulations with $\alpha=6.49\cdot  10^{-4}$ and $\alpha'=4.635\cdot  10^{-4}$ (corresponding to  $T=0.8\,$K, see Table \,\ref{t:1}), shown by solid blue lines,  and with the same  $\alpha=6.49\cdot  10^{-4}$, but $\alpha'=0$ (dashed red lines).
}
\end{figure*}
\subsection{\label{ss:res0lution} Vortex tangles demonstrate weak Kelvin wave turbulence }

 The previous section ended with the observation that at $T=0\,$K, the Kelvin wave cascade indeed propagates almost up to the upper end of the available $k$ interval, while for $T\geq 0.8\,$K the cascade is only barely visible. This means that even very small mutual friction with  $\alpha\lesssim 10^{-3}$  effectively suppresses the cascade. This conclusion is in agreement with the analytical analysis in Ref.\,\cite{BLP12} according to which for  $\alpha \lesssim 10^{-2}$ the Kelvin wave cascade is fully damped.

  In our simulations, the line-smoothing procedure at the scale resolution serves as an effective mechanism for damping the Kelvin waves. Therefore, it is important to know how our results depend on the   space resolution. For this goal we  plot in  Figs.\,\ref{f:4}A-\ref{f:4}C the evolution of $\<\C L (t)/\C L_0\>$ for   $T=1.1\,,\ 0.8\,$K,  and $0\,$K with three  initial spatial resolutions:  $\Delta \xi \sb {mid}=5\cdot 10^{-4}\,$cm, as in all the previous plots, (green dash lines), a higher resolution, $\Delta \xi \sb{high}=3\cdot 10^{-4}\,$cm  (solid blue lines) and a lower resolution, $\Delta \xi \sb {low}=8\cdot 10^{-4}\,$cm (dash-dotted red lines). We see that for $T=1.1\,,\ 0.8\,$K, (Figs.\,\ref{f:4}A, \ref{f:4}B),  all the three lines practically coincide, as expected: at these temperatures the Kelvin wave cascade is fully damped at a wave length $\lambda\sim \ell \gg \Delta \xi$ and therefore the motion at scales about $\Delta \xi$ plays no role. Some minor dependence of the VLD evolution is observed at $T=0\,$K (Fig.\,\ref{f:4}C) when the Kelvin waves cascade does reach the resolution limit. This can be caused by several reasons, related to the finiteness of the scale separation between $\ell$ and $\Delta \xi$. Their analysis is beyond the scope of this paper.

   The evolution of $\ell\widetilde S(t)$, shown in Fig.\,\ref{f:5}A-\ref{f:5}C, for the same set of parameters is much more sensitive to the resolution. This is expected, because the curvature is determined by the shortest Kelvin waves in the system. Correspondingly, at the higher temperature $T=1.1\,$K (Fig.\,\ref{f:5}A) where the Kelvin waves are excited only with $\lambda \simeq \ell$, there is practically  no dependence on the resolution. For $T=0.8\,$K, when only the tails of the Kelvin waves energy spectrum reaches the resolution scale one sees in Fig.\,\ref{f:5}B some weak dependence on the resolution.

   According to our understanding,  the Kelvin waves cascade reaches the resolution limit $k\sb{max}  \lesssim k\sb{res}\simeq \pi /\Delta \xi$, only at ultra-low temperatures, below   $T=0.5\,$K.  This should lead to an essential dependence of $\widetilde S \ell$ on the resolution. This is indeed the case, as one sees in Fig.\,\ref{f:5}C.

 Employing the LN spectrum  of weak-wave turbulence\,\eqref{curvC} at $T=0$K we predicted  $\ell \tilde S \propto  \Delta \xi^{-2/3}$ according Eq.\,\eqref{LN}.   Thus the curvature $\ell \widetilde S$   calculated with different line resolutions and  compensated by $(\Delta \xi/\Delta \xi \sb {mid})^{2/3} $ should collapse   to a single line. This is indeed the case,  cf. Fig.\,\ref{f:6}A in comparison with the non-compensated curvatures in  Fig.\,\ref{f:5}C. For completeness we show in Fig.\,\ref{f:6}B the time dependence of the dimensionless curvature $\ell \widetilde S$, compensated by $(\Delta \xi/\Delta \xi \sb {mid})^{4/5}$, as predicted by Eq.\,\eqref{curvE} for the KS-spectrum\,\eqref{KS} of weak wave turbulence. According to our understanding,  collapse in Fig.\,\ref{f:6}A  is slightly better than in  Fig.\,\ref{f:6}B, at least for later times, $t>3\,$s, when a significant decay  of the vortex line density occurs, see Fig.\,\ref{f:2}A. Nevertheless the difference   between these two cases (LN  and KS-spectra of weak wave turbulence) is modest and cannot serve as a decisive argument.   This is not our goal here. As we explained in the Introduction this can be done by specially designed numerical simulations like those in Refs.\,\cite{Krs12,BL13}.

 At this point, we can pose the important question of the nature of the Kelvin wave turbulence. Is it the weak turbulence characterized by the one of the two spectra (LN or KS)   or the strong turbulence with the Vinen spectrum\,\eqref{V}, for which the data should collapse with the linear compensation by $ \Delta \xi/\Delta \xi \sb {mid}$, shown in Fig.\,\ref{f:6}C.
One sees that the lines do not  collapse, in contrast to  Figs.\,\ref{f:6}A and \ref{f:6}B with the LN  and KS compensation. We consider these results as a clear evidence of a weak wave turbulence regime (presumably with the LN spectrum) in the decay of counterflow turbulence at ultralow temperatures.

 There exist analytical arguments in favor of weak wave turbulence regime of energy cascade by Kelvin waves. Using LN kinetic equation\,(5) suggested in Ref.\,\cite{LN1}, the  damping frequency of Kelvin waves can be estimated as:
 \begin{subequations}\label{comp}
 \begin{equation}\label{gamm}
 \gamma(k)\simeq \frac{\varepsilon ^{2/3}}{\kappa} k^{2/3}\ell^{4/3}\,,
 \end{equation}
and compared with the Kelvin wave frequency
\begin{equation}\label{gamma}
\omega(k)= \frac{\Lambda \kappa}{4\pi} k^2\simeq \kappa k^2\ .
 \end{equation}
  \end{subequations}
  The last estimate accounts for that numerically $\Lambda/(4\pi)$ is close to unity.
  It is known\,\cite{ZLF} that the applicability criterion of the regime of weak-wave turbulence requires $\gamma(k)\ll \omega(k)$.  In our case,
    \begin{subequations}\label{comp}
 \begin{equation}\label{zeta1}
    \zeta(k)\equiv \frac{\gamma(k)}{\omega(k)}
    \simeq  \frac{\varepsilon ^{2/3}\, \ell^{4/3}}{\kappa^2 \, k^{4/3}} \ .
 \end{equation}
 With the estimate\,\eqref{nup} for the rate of energy dissipation $\varepsilon$ via effective Vinen's viscosity $\nu'$, Eq. \eqref{zeta1} gives:
 \begin{equation}\label{zeta2}
    \zeta(k)
    \simeq  \Big (\frac{\nu'}{\kappa} \Big )^{2/3}  \frac 1 {(k\ell)^{4/3}} \ .
 \end{equation}
 \end{subequations}
 In our simulations and experimentally\,\cite{WG} $\nu'\simeq 0.1\,\kappa$. This means that even at the beginning of the inertial interval, when $k\ell\simeq 1$, $\zeta <1$ and one should expect a regime which is close to the weak-wave turbulence case. As $k$ increases, $\zeta$ quickly decreases and becomes much smaller than unity at the short-wave length edge of the inertial interval, where $k\sb{max}\simeq 1/(\Delta \xi)$. Here we expect a pure weak-wave turbulence regime, presumably with the LN spectrum.

 Notice, that during the decay of the vortex tangle, the total width of the inertial interval $\ell/(\Delta\xi)$ increases and therefore the approximation of the weak wave turbulence becomes better. In particular, this explains why the LN collapse in Fig.\,\ref{f:6} becomes better at larger times, $t>3$\,s.

 Notice that the collapse Eqs.\,\eqref{Curv} assume that the parameter $\Phi$, defined by Eq.\,\eqref{curvC1},  is independent of the resolution $\Delta\xi$.  This is a reasonable approximation because the integral in Eq.\,\eqref{curvC1} is  dominated by the lower cutoff of the power spectra, $k\sb{min}$.  For example, for the LN spectrum\,\eqref{LN} the  $\Phi(\Delta\xi)$ dependence can be estimated as $\Phi(\Delta\xi)\propto \big[1- \frac 12 (\Delta \xi/\ell)^{2/3}\big]$, and consequently  can be approximated by a  constant with accuracy  better than  6\% for our numerical parameters even at small times $t<1\,$s. During the time evolution $\ell$ increases and the accuracy improves further.

 We also have to mention that the difference in the  cutoff lengths for low  and  high resolution is only a factor of $\frac83\approx 2.7$ and cannot be straightforwardly  used (for example by plotting $ \ell \widetilde S $  versus $\ell k\sb{max}$) for any kind of argument in favor of power law scaling. Instead, we have plotted in Fig.\,\ref{f:6} the scale compensated plots $ \ell \widetilde S (\ell k\sb{max})^x$ versus $t$ and see a clear difference for $x=\frac23,\, \frac45$ and 1  in Figs.~\ref{f:6}A, \ref{f:6}B and \ref{f:6}C.

 Definitely,  it would be nice to see some Kelvin waves spectra at low temperatures, for example, analyzing a configuration like the one observed in Fig.\,\ref{f:1}E, where a long filament with Kelvin waves is visible.  Unfortunately, this is an almost impossible task
with our numerical resolutions (with only about 25--30 points on the intervortex scale $\ell$), giving less than a decade of inertial interval with Kelvin waves.  As we mentioned in the Introduction, it is hardly possible to distinguish between the theoretical predictions\,\eqref{V}, \eqref{KS}, and \eqref{LN},  even with the better resolution achievable  in studies of vortex line evolutions from much simpler initial configurations in Refs.\,\cite{AT00,KVSB01}. This can be done only with specially arranged simulations in which Kelvin waves are excited on a straight line, say with 1024 points, as in Ref.\,\cite{BL13}.

 Our goal in this paper was different: to demonstrate that Kelvin waves play a statistically important role in the decay of a random vortex tangle,  generated by  counterflow turbulence. Fortunately, analyzing  the time evolution of the
mean re-scaled vortex curvature in Figs.\,\ref{f:6} (instead of the direct measurement  of
the scaling exponents) we were able to distinguish between strong and weak wave turbulence spectra of Kelvin waves. In some sense, this approach is an analog of the extended self-similarity analysis by Benzi et al\,\cite{ESS}, successfully
used in the past for the scaling analysis of the experimental and
numerical data in developed classical turbulence.

\section*{\label{S:sum}Summary}
\begin{itemize}
\item We reported a comprehensive numerical study of the statistical properties of the free decaying vortex tangle at low and ultra-low temperatures. The simulations were carried out in the framework of the full Biot-Savart equations using vortex filament method with high spatial resolution $\Delta \xi$ which encompasses two decades of the Kelvin wave cascade below the intervortex scale $\ell$.

 \item We   estimated   from our numerical results the effective Vinen's viscosity and demonstrated its good agreement with the experimental results\,\cite{WG} in Manchester spin-down experiment in $^4$He.

\item We showed that even very small mutual friction with $\alpha  \gtrsim  10^{-4}$ effectively suppresses the Kelvin wave cascade such that there survive only Kelvin waves with the wavelength $\lambda \simeq \ell$. At ultra-low temperatures,  $T \lesssim 0.5\,$K, when $\alpha   \lesssim 10^{-5}$, the Kelvin wave cascade is developing up to the resolution scale with $\lambda \simeq  2 \cdot \Delta \xi$, where it is terminated  in the simulations by the line-smoothing procedure that mimics the  dissipation of Kelvin waves by the emission of phonons and rotons at the vortex-core scale in real flows.

\item We showed that the Kelvin wave cascade is statistically important: the shortest available Kelvin waves at the end of the cascade determine the mean vortex line curvature $\widetilde S$, giving $\widetilde S \simeq  30/\ell$.

\item   We showed that in the decay of counterflow turbulence there is regime of weak turbulence of Kelvin waves with the LN spectrum\,\eqref{LN} rather than the strong turbulence regime with the Vinen spectrum\,\eqref{V}.

 \end{itemize}

\acknowledgements
 We acknowledge useful comments by Andrei Golov and all the three anonymous referees.
This paper had been supported in part by the Minerva
Foundation, Munich, Germany and by Grant 13-08-00673
from RFBR (Russian Foundation of Fundamental Research).
L.K. acknowledges the kind hospitality at the Weizmann
Institute of Science during the main part of the project.\\~\\

\appendix
\subsection*{\label{a:1}Appendix.  Importance of the energy-conserving mutual friction force}
As we discussed in Sec.\,\ref{s:num},   the experimental data for the mutual friction parameters are not available for temperatures below $1\,$K. Parameter $\alpha$ can be estimated analytically by Eq.\,\eqref{alpha}. Corresponding equation for $\alpha'$ is not available. One may think that very small $\alpha'$, responsible for non-dissipative interactions is not important and can be neglected. To check this idea, we compare in Fig.\,\ref{f:7}  numerical results for the VLD and curvature evolution of the tangle with  $\alpha=6.49\cdot  10^{-4}$ and $\alpha'=4.635\cdot  10^{-4}$,  corresponding to  $T=0.8\,$K,  with the evolution with the same $\alpha$, but $\alpha'=0$. One sees that the results differ essentially and conclude that even very small $\alpha'$ cannot be neglected. This can be related to the geometry of the   vortex lines approaching the reconnection, that effect the reconnection rate.

\end{document}